\documentclass[twocolumn,journal,10pt]{IEEEtran}
\usepackage[T1]{fontenc}
\usepackage[latin9]{inputenc}
\usepackage{xcolor}
\usepackage{float}
\usepackage{amsmath}
\usepackage{amsthm}
\usepackage{amssymb}
\usepackage{graphicx}
\usepackage[unicode=true,
 bookmarks=true,bookmarksnumbered=true,bookmarksopen=true,bookmarksopenlevel=1,
 breaklinks=false,pdfborder={0 0 0},pdfborderstyle={},backref=false,colorlinks=false]
 {hyperref}
\hypersetup{pdftitle={Your Title},
 pdfauthor={Your Name},
 pdfpagelayout=OneColumn, pdfnewwindow=true, pdfstartview=XYZ, plainpages=false}

\makeatletter
\theoremstyle{plain}
\newtheorem{thm}{\protect\theoremname}
\theoremstyle{definition}
\newtheorem{defn}[thm]{\protect\definitionname}
\theoremstyle{plain}
\newtheorem{lem}[thm]{\protect\lemmaname}
\theoremstyle{remark}
\newtheorem{rem}[thm]{\protect\remarkname}

\usepackage[caption=false,font=footnotesize]{subfig}

\usepackage[caption=false,font=footnotesize]{subfig}

\usepackage{amstext}
\usepackage{pdfsync}
\usepackage{amsthm}

\usepackage{epstopdf}

\usepackage{tikz}
\usetikzlibrary{arrows}
\usetikzlibrary{shadows}
\tikzset{
  every overlay node/.style={
    draw=white,anchor=north west,
  },
}
\usetikzlibrary{positioning, decorations.pathmorphing}

\usepackage{hyperref}
\@ifundefined{rangeHsb}{\usepackage{xcolor}}{}

\usepackage{algorithm,algpseudocode}

\usepackage{lipsum}

\makeatother

\providecommand{\definitionname}{Definition}
\providecommand{\lemmaname}{Lemma}
\providecommand{\remarkname}{Remark}
\providecommand{\theoremname}{Theorem}

\begin{document}
\title{Event-triggered Consensus of Matrix-weighted Networks Subject to Actuator
Saturation}
\author{Lulu~Pan,~Haibin~Shao,~\IEEEmembership{Member,~IEEE,} Yuanlong
Li, Dewei~Li,~and~Yugeng~Xi,~\IEEEmembership{Senior Member,~IEEE}\thanks{This work is supported by the National Science Foundation of China
(Grant No. 61973214, 62022055, 61590924, 61963030) and Natural Science
Foundation of Shanghai (Grant No. 19ZR1476200). \textcolor{black}{(Corr}esponding
author: Haibin Shao)}\thanks{Lulu Pan, Haibin Shao, Yuanlong Li, Dewei Li and Yugeng Xi are with
the Department of Automation and the Key Laboratory of System Control
and Information Processing Ministry of Education of China, Shanghai
Jiao Tong University, Shanghai 200240, China (\{llpan,shore,liyuanlong0301,dwli,ygxi\}@sjtu.edu.cn). }}
\maketitle
\begin{abstract}
The ubiquitous interdependencies among higher-dimensional states of
neighboring agents can be characterized by matrix-weighted networks.
This paper examines event-triggered global consensus of matrix-weighted
networks subject to actuator saturation. Specifically, a distributed
dynamic event-triggered coordination strategy, whose design involves
sampled state of agents, saturation constraint and auxiliary systems,
is proposed for this category of generalized network to guarantee
its global consensus. Under the proposed event-triggered coordination
strategy, sufficient conditions are derived to guarantee the leaderless
and leader-follower global consensus of the multi-agent systems on
matrix-weighted networks, respectively. The Zeno phenomenon can be
excluded for both cases under the proposed coordination strategy.
It turns out that the spectral properties of matrix-valued weights
are crucial in event-triggered mechanism design for matrix-weighted
networks with actuator saturation constraint. Finally, simulations
are provided to demonstrate the effectiveness of proposed event-triggered
coordination strategy. This work provides a more general design framework
compared with existing results that are only applicable to scalar-weighted
networks.
\end{abstract}


\IEEEpeerreviewmaketitle{}

\section{Introduction}

\IEEEPARstart{C}{onsensus} problem on matrix-weighted networks is
becoming a recent concern since, as an immediate generalization of
scalar-weighted networks, they naturally captures interdependencies
among higher-dimensional states of neighboring agents in a multi-agent
network \cite{sun2018dimensional,pan2021bipartite,pan2018bipartite,trinh2018matrix,mesbahi2010graph,olfati2007consensus}.
Actually, matrix-weighted networks arise in scenarios such as graph
effective resistance based distributed control and estimation \cite{barooah2006graph,tuna2017observability},
logical inter-dependency of multiple topics in opinion evolution \cite{friedkin2016network},
bearing-based formation control \cite{zhao2015translational}, array
of coupled LC oscillators \cite{tuna2016synchronization} as well
as consensus and synchronization on matrix-weighted networks \cite{pan2018bipartite,trinh2018matrix}.

In contrast to scalar-weighted networks, connectivity alone does not
translate to achieving consensus on matrix-weighted networks, properties
of weight matrices now play an important role in the characterization
of consensus as well as the design of interaction protocols subject
to physical constraints. In literatures, positive/negative definiteness/semi-definiteness
weight matrices have been employed to provide consensus conditions
\cite{pan2018bipartite,su2019bipartite,pan2021bipartite,trinh2018matrix}.
In the meantime, it is worth noting that the matrix-weighted network
is a more general category of multi-agent networks, recent trends
in this line of research seems to involve the constraints of physical
systems encountered in real-world applications. For instance, beyond
the first-order local dynamics, consensus conditions for second-order
multi-agent system on matrix-weighted networks are provided \cite{miao2021second,wang2020consensus}.
However, a comprehensive investigation of matrix-weighted networks
subject to physical constraints is still lacking. Typically, these
constraints can arise from input, output, and communication, which
bring nonlinearities in the closed-loop dynamics \cite{li2018stability,hu2001control,sussmann1993general,wang2010event,aastrom1999comparison}. 

For distributed control of practical multi-agent systems, the control
input is often subject to saturation constraint due to physical limitations.
In literatures, insightful efforts have been devoted to cooperative
control of multi-agent systems subject to input saturation via continuous-time
information exchange. For instance, global consensus problem of single-integrator
and double-integrator multi-agent systems with input saturation were
examined in \cite{li2011consensus,meng2013global}. Moreover, it was
shown in \cite{meng2013global,yang2014global} that global leader-following
consensus of neutrally stable linear multi-agent systems with input
saturation can be achieved using linear local feedback laws. By using
the low-gain feedback design technique, semi-global consensus can
be achieved for linear multi-agent systems with input saturation whose
open-loop poles are all located in the closed left-half plane \cite{su2013semi,lin1999low}.

However, in the aforementioned investigations, the simultaneous information
exchange and transmission between neighboring agents are needed, which
is expensive from the perspective of both communication and computation.
The event-triggered mechanism turns out to be efficient in handling
this issue, where the control actuation or the information transmission
was determined by the designed event \cite{ding2017overview,nowzari2019event}.
A decentralized event-triggered control for single-integrator multi-agent
systems was initially proposed in \cite{dimarogonas2011distributed}
where the event-triggered function for the agent depends on the continuous
information monitoring of its neighbors. In order to overcome this
limitation, the distributed event-triggered functions proposed in
\cite{nowzari2016distributed} where only state of neighboring agents
at last event-triggered time was employed to avoid the continuous
information exchange between neighboring agents.  However, this method
was not satisfactory in the respect of avoiding the Zeno behaviors.
In \cite{yi2018dynamic}, distributed event-triggered consensus control
of single-integrator multi-agent systems was examined and it was shown
that dynamic parameters ensures less triggering instants and played
essential roles in avoiding Zeno behaviors. For more details about
event-triggered problem of multi-agent systems, one can refer to the
recent survey papers \cite{ding2017overview,nowzari2019event}. 

In contrast to the numerous results on multi-agent systems with saturated
control or event-triggered control, very little attention is spent
on the consensus problem of multi-agent systems under the constraints
of both input saturation and event-triggered communication. In this
line of works, the influence of actuator saturation on event-triggered
control for single systems is examined in \cite{kiener2014actuator}.
In \cite{wu2016distributed}, a distributed event-triggered control
strategy is proposed to achieve consensus for multi-agent systems
subject to input saturation through output feedback, however the Zeno
behavior therein cannot be avoided. In \cite{yin2016adaptive}, LMI
techniques are employed to design leader-following consensus protocol
for multi-agent systems subject to input saturation, but the design
depends on the global information of graph Laplacian. Recently, the
event-triggered global consensus problem for leaderless multi-agent
systems with input saturation constraints using a triggering function
whose threshold depends on time rather than state is studied \cite{yi2019distributed}. 

Although the event-triggered consensus problem with input saturation
constraint for scalar-weighted networks has been investigated, it
turns out that the existing methods are only applicable to scalar-weighted
networks. For the case of matrix-weighted networks, it becomes more
challenging since specific properties of weight matrices have to be
involved to the design of the interaction protocol for multi-agent
networks under the constraints of input saturation and event-triggered
communication, which makes this work non-trivial. To the best of our
knowledge, this paper is the first attempt to examine the interaction
protocol design problem for matrix-weighted networks subject to both
actuator saturation and event-triggered communication. 

The main contributions of this paper are as follows. A novel distributed
event-triggered coordination strategy with dynamic parameters in triggering
function design are introduced for multi-agent system on matrix-weighted
networks subject to actuator saturation. The update of dynamic parameter
for each agent is determined by the measurement error, the saturated
state difference between each agent and its neighbors at triggering
instants, as well as the largest eigenvalue of local accessible matrix-valued
edge weights. The continuous state exchange between neighboring agents
can be avoided in our design. Sufficient conditions are derived to
guarantee the global bipartite consensus for both leaderless and leader-follower
multi-agent system on matrix-weighted networks subject to actuator
saturation. In the meantime, it is shown that the Zeno phenomenon
for both cases can be avoided. Moreover, the proposed event-triggered
coordination strategy involving actuator saturation constraint in
this work provides a more general design framework compared with existing
results that are only applicable to scalar-weighted networks \cite{yi2019distributed}.

The remainder of this paper is organized as follows. The preliminaries
of matrix analysis and graph theory are introduced in \S 2 as well
as fundamental facts of matrix-weighted networks. Then, the problem
formulation is provided in \S 3 and the main results on the design
of event-triggered bipartite consensus protocol for leaderless matrix-weighted
networks and leader-follower matrix-weighted networks are provided
in \S 4 and \S 5, respectively, which is followed by the numerical
simulation in \S 6. The concludin\textcolor{black}{g remarks are
finally given in \S 7.}

\section{Preliminaries}

In this section, we provide notations and background knowledge of
matrix-weighted networks.

\subsection{Notations}

\textcolor{black}{Let $\mathbb{R}$ and $\mathbb{Z}_{+}$ be the set
of real numbers and positive integers, respectively. }Denote $\underline{n}=\left\{ 1,2,\ldots,n\right\} $
for a $n\in\mathbb{Z}_{+}$. A symmetric matrix $M\in\mathbb{R}^{n\times n}$
is positive definite (Resp. negative definite), denoted by $M>0$
(Resp. $M<0$), if $\boldsymbol{z}^{T}M\boldsymbol{z}>0$ (Resp. $\boldsymbol{z}^{T}M\boldsymbol{z}<0$)
for all $\boldsymbol{z}\in\mathbb{\mathbb{R}}^{n}$ and $\boldsymbol{z\not}=\boldsymbol{0}$
and is positive (Resp. negative) semi-definite, denoted by $M\ge0$
(Resp. $M\le0$), if $\boldsymbol{z}^{T}M\boldsymbol{z}\ge0$ ($\boldsymbol{z}^{T}M\boldsymbol{z}\le0$)
for all $\boldsymbol{z}\in\mathbb{\mathbb{R}}^{n}$. The absolute
value of a\textcolor{black}{{} symmetric matrix $M\in\mathbb{R}^{n\times n}$
is denoted by $|M|$ such that $|M|=M$ if $M>0$ or $M\ge0$ and
$|M|=-M$ if $M<0$ or $M\le0$. }The absolute value of a\textcolor{black}{{}
vector $\boldsymbol{z}=(z_{1},z_{2},\cdots,z_{n})^{T}\in\mathbb{R}^{n}$
is denoted by $|\boldsymbol{z}|$ such that $|\boldsymbol{z}|=(|z_{1}|,|z_{2}|,\cdots,|z_{n}|)^{T}$.
Denote by $\boldsymbol{z}>0$ if $\boldsymbol{z}=|\boldsymbol{z}|$
and $\boldsymbol{z\not}=\boldsymbol{0}$. The null space of a matrix
$M\in\mathbb{R}^{n\times n}$ is $\text{{\bf null}}(M)=\left\{ \boldsymbol{z}\in\mathbb{R}^{n}|M\boldsymbol{z}=\boldsymbol{0}\right\} $.
Let $\lambda_{n}(M)$ denote the largest eigenvalue of a symmetric
matrix $M\in\mathbb{R}^{n\times n}$. $\boldsymbol{1_{n}}\in\mathbb{R}^{n}$
and $0_{n\times n}\in\mathbb{R}^{n\times n}$ designate the vector
whose components are all $1$'s and the matrix whose components are
all $0$'s, respectively. The sign function $\text{{\bf sgn}}(\cdot):\mathbb{R}^{n\times n}\mapsto\left\{ 0,-1,1\right\} $
satisfies $\text{{\bf sgn}}(M)=1$ if $M\geq0$ or $M>0$, $\text{{\bf sgn}}(M)=-1$
if $M\leq0$ or $M<0$, and $\text{{\bf sgn}}(M)=0$ if $M=0_{n\times n}$.
}\textcolor{magenta}{}

\subsection{Matrix-weighted Networks}

Let $\mathcal{G}=(\mathcal{V},\mathcal{E},A)$ be a matrix-weighted
network where the node set and the edge set of $\mathcal{G}$ are
denoted by $\mathcal{V}=\left\{ 1,2,\ldots,n\right\} $ and $\mathcal{E}\subseteq\mathcal{V}\times\mathcal{V}$,
respectively. The matrix weight for edges in $\mathcal{G}$ is a symmetric
matrix $A_{ij}\in\mathbb{R}^{d\times d}$ such that $|A_{ij}|\geq0$
or $|A_{ij}|>0$ if $(i,j)\in\mathcal{E}$ and $A_{ij}=0_{d\times d}$
otherwise for all $i,j\in\mathcal{V}$. Thereby, the matrix-valued
adjacency matrix $A=[A_{ij}]\in\mathbb{R}^{dn\times dn}$ is a block
matrix such that the block located in the $i$-th row and the $j$-th
column is $A_{ij}$. We shall assume that $A_{ij}=A_{ji}$ for all
$i\not\not=j\in\mathcal{V}$ and $A_{ii}=0_{d\times d}$ for all $i\in\mathcal{V}$,
which are analogous to the assumptions of undirected and simple graph
in a normal sense. The neighbor set of an agent $i\in\mathcal{V}$
is denoted by $\mathcal{N}_{i}=\left\{ j\in\mathcal{V}\,|\,(i,j)\in\mathcal{E}\right\} $.
Denote $D=\text{{\bf diag}}\left\{ D_{1},D_{1},\cdots,D_{n}\right\} \in\mathbb{R}^{dn\times dn}$
as the matrix-weighted degree matrix of a graph where $D_{i}=\sum_{j\in\mathcal{N}_{i}}|A_{ij}|\in\mathbb{R}^{d\times d}$.
The matrix-valued Laplacian matrix of a matrix-weighted graph is defined
as $L(\mathcal{G})=D-A$.\textcolor{pink}{{} }
\begin{defn}
A bipartition of node set $\mathcal{V}$ of matrix-weighted network
$\mathcal{G}=(\mathcal{V},\mathcal{E},A)$ is two subsets of nodes
$\mathcal{V}_{i}\subset\mathcal{V}$, where $i\in\underline{2}$,
such that $\mathcal{V}=\mathcal{V}_{1}\cup\mathcal{V}_{2}$ and $\mathcal{V}_{1}\cap\mathcal{V}_{2}=\textrm{Ø}$. 
\end{defn}

In signed networks, the concept of structural balance (can be tracked
back to the seminal work\ \cite{harary1953notion}) turns out to
be an important graph-theoretic object playing a critical role in
bipartite consensus problems \cite{altafini2013consensus}. This concept
has been extended to the matrix-weighted networks in \cite{pan2018bipartite}.
\begin{defn}
\cite{pan2018bipartite}\label{def:SB-weighted-network} A matrix-weighted
network $\mathcal{G}=(\mathcal{V},\mathcal{E},A)$ is structurally
balanced if there exists a bipartition of the node set $\mathcal{V}$,
say $\mathcal{V}_{1}$ and $\mathcal{V}_{2}$, such that the matrix
weights on the edges within each subset is positive definite or positive
semi-definite, but negative definite or negative semi-definite for
the edges between the two subsets. A matrix-weighted network is structurally
imbalanced if it is not structurally balanced.
\end{defn}
Let $\mathcal{G}=(\mathcal{V},\mathcal{E},A)$ be a matrix-weighted
network with a node bipartition $\mathcal{V}_{1}$ and $\mathcal{V}_{2}$
and $d\in\mathbb{N}$ represent the dimension of edge weight. The
gauge transformation for this node bipartition $\mathcal{V}_{1}$
and $\mathcal{V}_{2}$ is performed by a diagonal matrix $D^{*}=\text{{\bf diag}}\left\{ \sigma_{1},\sigma_{2},\ldots,\sigma_{n}\right\} $
where $\sigma_{i}=I_{d}$ if $i\in\mathcal{V}_{1}$ and $\sigma_{i}=-I_{d}$
if $i\in\mathcal{V}_{2}$. If the matrix-weighted network $\mathcal{G}=(\mathcal{V},\mathcal{E},A)$
is structurally balanced, then it satisfies that $D^{*}AD^{*}=[|A_{ij}|]\in\mathbb{R}^{dn\times dn}$. 

The following result characterizes the structure of the null space
of matrix-valued Laplacian for matrix-weighted networks, which is
different from the Laplacian matrix for scalar-weighted networks where
the null space of the Laplacian matrix is $\text{{\bf span}}\left\{ \boldsymbol{1}_{dn}\right\} $.
\begin{lem}
\label{lem:1-1}\cite{pan2018bipartite} Let $\mathcal{G}=(\mathcal{V},\mathcal{E},A)$
be a structurally balanced matrix-weighted network. Then the Laplacian
matrix $L$ of $\mathcal{G}$ is positive semi-definite and its null
space can be characterized by $\text{{\bf null}}(L)=\text{{\bf span}}\left\{ \mathcal{R},\mathcal{H}\right\} ,$
where 
\[
\mathcal{R}=\text{{\bf range}}\{D^{*}(\boldsymbol{1}_{n}\otimes I_{d})\}
\]
and
\begin{align*}
\mathcal{H=}\{\boldsymbol{v} & =(\boldsymbol{v}_{1}^{T},\boldsymbol{v}_{2}^{T},\cdots,\boldsymbol{v}_{n}^{T})^{T}\in\mathbb{R}^{dn}\mid\\
 & (\boldsymbol{v}_{i}-\text{{\bf sgn}}(A_{ij})\boldsymbol{v}_{j})\in\text{{\bf null}}(|A_{ij}|),\,(i,j)\in\mathcal{E}\}.
\end{align*}
\end{lem}

\section{Problem Formulation}

Consider a multi-agent system on matrix-weighted network $\mathcal{G}=(\mathcal{V},\mathcal{E},A)$
with $n\in\mathbb{Z}_{+}$ agents, the dynamics of the $i$th agent
reads,
\begin{equation}
\dot{\boldsymbol{x}}_{i}(t)=\text{{\bf sat}}_{\varDelta}(\boldsymbol{u}_{i}(t)),i\in\mathcal{V},\label{eq:the agent protocol}
\end{equation}
where $\boldsymbol{x}_{i}(t)\in\mathbb{R}^{d}$ and $\boldsymbol{u}_{i}(t)\in\mathbb{R}^{d}$
are the state and control input associated with agent $i$. For a
given saturation level $\Delta>0$, $\text{{\bf sat}}_{\varDelta}:\mathbb{R}\mapsto\mathbb{\mathbb{R}}$
denote the saturation function such that\textcolor{black}{
\[
\text{{\bf sat}}_{\varDelta}(h_{i})=\text{{\bf sgn}}(h_{i})\text{{\bf min}}\left\{ |h_{i}|,\Delta\right\} ,i\in\underline{l}.
\]
and
\[
\text{{\bf sat}}_{\varDelta}(\boldsymbol{h})=(\text{{\bf sat}}_{\varDelta}(h_{1}),\text{{\bf sat}}_{\varDelta}(h_{2}),\cdots,\text{{\bf sat}}_{\varDelta}(h_{l}))^{T},
\]
where $\boldsymbol{h}=(h_{1},h_{2},\cdots,h_{l})^{T}\in\mathbb{R}^{l}$
and $l\in\mathbb{Z}_{+}$.} One can conclude the following facts on
saturation function, which is crucial in the subsequent theoretical
analysis.
\begin{lem}
\textcolor{black}{\label{saturation-inequality}For any $\boldsymbol{h}=(h_{1},h_{2},\cdots,h_{l})^{T}\in\mathbb{R}^{l}$
where }$l\in\mathbb{Z}_{+}$, the following inequality holds 
\[
\text{{\bf sat}}_{\varDelta}\left(\boldsymbol{h}\right)^{T}\text{{\bf sat}}_{\varDelta}\left(\boldsymbol{h}\right)\leq\boldsymbol{h}^{T}\text{{\bf sat}}_{\varDelta}\left(\boldsymbol{h}\right).
\]
\end{lem}
In the following discussion, we proceed to design control law design
for multi-agent system \textcolor{black}{\eqref{eq:the agent protocol}}
on matrix-weighted networks such that global bipartite consensus can
be guaranteed without continuous state information exchange amongst
agents. We shall first examine matrix-weighted networks without leaders,
namely, leaderless matrix-weighted networks.

\section{Leaderless Matrix-weighted Networks}

\subsection{Actuator Saturation}

In the following discussions, we assume that the Laplacian matrix
$L$ corresponding to the matrix-weighted network $\mathcal{G}=(\mathcal{V},\mathcal{E},A)$
satisfies the following assumption.

\textbf{Assumption} 1\label{null-space-SB}. There exists a gauge
transformation $D^{*}$ such that $\text{{\bf null}}(D^{*}LD^{*})=\mathcal{R}$.

In this section, before we give the event-triggered coordination strategy
for the multi-agent systems \eqref{eq:the agent protocol} on matrix-weighted
networks, we shall first discuss whether the multi-agent system \eqref{eq:the agent protocol}
on matrix-weighted networks can achieve the global bipartite consensus
only under saturated control protocol. Consider the following distributed
continuous-time protocol,

\textcolor{black}{
\begin{equation}
\boldsymbol{u}_{i}(t)=-\sum_{j\in\mathcal{N}_{i}}|A_{ij}|(\boldsymbol{x}_{i}(t)-\text{{\bf sgn}}(A_{ij})\boldsymbol{x}_{j}(t)),i\in\mathcal{V},\label{eq:matrix-bipartite-consensus-for-agent}
\end{equation}
the overall dynamics of the multi-agent system \eqref{eq:the agent protocol}
can be characterized by the associated matrix-valued Laplacian,
\begin{equation}
\dot{\boldsymbol{x}}(t)=\text{{\bf sat}}_{\varDelta}(-L\boldsymbol{x}(t)),\label{equ:matrix-consensus-overall}
\end{equation}
where $\boldsymbol{x}(t)=(\boldsymbol{x}_{1}^{T}(t),\boldsymbol{x}_{2}^{T}(t),\ldots,\boldsymbol{x}_{n}^{T}(t))^{T}\in\mathbb{R}^{dn}$. }
\begin{defn}
\textcolor{black}{For each agent $i\in\mathcal{V}$ and an arbitrary
$\boldsymbol{x_{i}}(0){\color{red}{\color{black}\in\mathbb{R}^{d}}}$,
the multi-agent system \eqref{eq:the agent protocol} is said to admit
global bipartite consensus if ${\color{black}{\color{blue}{\color{black}\lim{}_{t\rightarrow\infty}\mid\boldsymbol{x}_{i}(t)\mid=\boldsymbol{\alpha}}}}$
where $\boldsymbol{\alpha}{\color{red}{\color{black}\in\mathbb{R}^{d}}}$
and $\boldsymbol{\alpha}>0$.}
\end{defn}
\begin{lem}
\textcolor{black}{\label{thm:leaderless-no-event}Let Assumption 1
holds. Then, under the control law \eqref{eq:matrix-bipartite-consensus-for-agent},
the multi-agent system \eqref{eq:the agent protocol} on the matrix-weighted
network $\mathcal{G}=(\mathcal{V},\mathcal{E},A)$ achieves global
bipartite consensus.}
\end{lem}
\begin{IEEEproof}
\textcolor{black}{Consider the following Lyapunov function candidate,}

\textcolor{black}{
\[
V(t)=\boldsymbol{x}^{T}(t)L\boldsymbol{x}(t),
\]
computing the time derivative of $V(t)$ along with \eqref{equ:matrix-consensus-overall}
yields,}

\textcolor{black}{
\begin{eqnarray*}
\dot{V}(t) & = & 2\boldsymbol{x}(t)^{T}L\dot{\boldsymbol{x}}(t)\\
 & = & \boldsymbol{x}(t)^{T}L\text{{\bf sat}}_{\varDelta}(-L\boldsymbol{x}(t)))\\
 & \le & 0.
\end{eqnarray*}
It is obvious that $\dot{V}(t)=0$ if and only if $L\boldsymbol{x}(t)=\boldsymbol{0}$,
i.e., $\boldsymbol{x}_{i}(t)=\text{{\bf sgn}}(A_{ij})\boldsymbol{x}_{j}(t),\,\forall i,j\in\underline{n}$.
Thus according to LaSalle's invariance principle \cite{Khalil}, 
\[
\underset{t\rightarrow\infty}{\text{{\bf lim}}}\left(\boldsymbol{x}_{i}(t)-\text{{\bf sgn}}(A_{ij})\boldsymbol{x}_{j}(t)\right)=\boldsymbol{0},\,\forall i,j\in\underline{n}.
\]
That is, the multi-agent system \eqref{eq:the agent protocol} achieves
global bipartite consensus under the control law \eqref{eq:matrix-bipartite-consensus-for-agent}.}
\end{IEEEproof}
\begin{rem}
\textcolor{black}{If the positive number $\varDelta$ is large enough,
the effect of the saturation function on the multi-agent system \eqref{equ:matrix-consensus-overall}
will vanish, then the result in Lemma \ref{thm:leaderless-no-event}
is in accordance with the result showed in \cite{pan2018bipartite,trinh2018matrix}.
Actually, for the saturation case, due to $\dot{V}(t)\leq0$, we know
that the saturation is no longer effective after a finite time which
depends on the initial value of each agent, the saturation function
and the network topology. }
\end{rem}
\textcolor{black}{From the above analysis, one can see that, to implement
consensus protocol \eqref{equ:matrix-consensus-overall} with saturation,
continuous states from neighbors are needed. However, continuous communication
is impractical in physical applications. To avoid continuously sending
information among agents and updating controls, in the following,
we shall equip the consensus protocol \eqref{equ:matrix-consensus-overall}
with an event-triggered communication strategy, in this setting, the
control signal is only updated when the triggering condition is satisfied. }

\subsection{\textcolor{black}{Event-triggered Mechanism Design}}

\textcolor{black}{Denote by $\widehat{\boldsymbol{x}}_{i}(t)$ as
the last broadcast state of agent $i\in\mathcal{V}$ at any given
time $t\geq0$, consider the following protocol for leaderless multi-agent
system with input saturation and event-triggered constraint,}

\textcolor{black}{
\begin{equation}
\dot{\boldsymbol{x}}_{i}(t)=\text{{\bf sat}}_{\varDelta}(\widehat{\boldsymbol{u}}_{i}(t)),i\in\mathcal{V},\label{eq:agent-event-triggered-protocol}
\end{equation}
}and

\textcolor{black}{
\begin{equation}
\widehat{\boldsymbol{u}}_{i}(t)=-\sum_{j\in\mathcal{N}_{i}}|A_{ij}|(\widehat{\boldsymbol{x}}_{i}(t)-\text{{\bf sgn}}(A_{ij})\widehat{\boldsymbol{x}}_{j}(t)),\label{eq:event-triggered-control-law}
\end{equation}
let $\widehat{\boldsymbol{x}}(t)=[\widehat{\boldsymbol{x}}_{1}^{T}(t),\widehat{\boldsymbol{x}}_{2}^{T}(t),\ldots,\widehat{\boldsymbol{x}}_{n}^{T}(t)]^{T}\in\mathbb{R}^{dn}$,
then the system \eqref{eq:agent-event-triggered-protocol} under the
control law \eqref{eq:event-triggered-control-law} can be written
in a compact form as 
\begin{equation}
\dot{\boldsymbol{x}}(t)=\text{{\bf sat}}_{\varDelta}(-L\widehat{\boldsymbol{x}}(t)).\label{eq:overall-event-trigger}
\end{equation}
}

\textcolor{black}{Define the state-based measurement error between
the last broadcast state of agent $i\in\mathcal{V}$ and its current
state at time $t\geq0$ as
\begin{equation}
\boldsymbol{e}_{i}(t)=\widehat{\boldsymbol{x}}_{i}(t)-\boldsymbol{x}_{i}(t),
\end{equation}
then the system-wise measurement error is denoted by $\boldsymbol{e}(t)=[\boldsymbol{e}_{1}^{T}(t),\boldsymbol{e}_{2}^{T}(t),\ldots,\boldsymbol{e}_{n}^{T}(t)]^{T}$.
For each agent $i\in\mathcal{V}$, the triggering time sequence is
initiated from $t_{1}^{i}=0$ and subsequently determined by,}

\textcolor{black}{
\begin{align}
t_{k+1}^{i} & =\underset{r\geq t_{k}^{i}}{\text{{\bf max}}}\{r\thinspace|\thinspace\theta_{i}(\varpi_{i}\parallel\boldsymbol{e}_{i}(t)\parallel^{2}-\rho_{i}\widehat{\boldsymbol{u}}_{i}^{T}(t)\text{{\bf sat}}_{\varDelta}\left(\widehat{\boldsymbol{u}}_{i}(t)\right))\nonumber \\
 & \leq\psi_{i}(t),\,\forall t\in[t_{k}^{i},r]\},\label{eq:event-triggered-time}
\end{align}
where $k\in\mathbb{Z}_{+}$, $\rho_{i}\in[0,1)$, $\theta_{i}$ and
$\varpi_{i}$ are the design parameters and $\psi_{i}(t)$ is an auxiliary
system for each agent $i\in\mathcal{V}$ such that,
\begin{eqnarray}
\dot{\psi}_{i}(t) & = & -\beta_{i}\psi_{i}(t)+\delta_{i}(\rho_{i}\widehat{\boldsymbol{u}}_{i}^{T}(t)\text{{\bf sat}}_{\varDelta}\left(\widehat{\boldsymbol{u}}_{i}(t)\right)\nonumber \\
 &  & -\varpi_{i}\parallel\boldsymbol{e}_{i}(t)\parallel^{2}),\label{eq:dynamic-function-autonomous}
\end{eqnarray}
with $\psi_{i}(0)>0$, $\beta_{i}>0$ and $\delta_{i}\in[0,1]$.}
\begin{thm}
\textcolor{black}{\label{event-triggered-leaderless}Consider the
multi-agent system \eqref{eq:overall-event-trigger} under the matrix-weighted
network $\mathcal{G}=(\mathcal{V},\mathcal{E},A)$ satisfying Assumptions
1. Let $\theta_{i}$ and $\varpi_{i}$ be such that $\theta_{i}>\frac{1-\delta_{i}}{\beta_{i}}$
and 
\begin{eqnarray*}
\varpi_{i} & =n\left({\displaystyle \sum_{j\in\mathcal{N}_{i}}}\lambda_{d}(\mid A_{ij}\mid)\right)^{2}+ & n\sum_{j\in\mathcal{N}_{i}}\lambda_{d}^{2}\left(\mid A_{ij}\mid\right),
\end{eqnarray*}
for all $i\in\mathcal{V}$, respectively, the triggering time sequence
is determined by \eqref{eq:event-triggered-time} for agent $i$ with
$\psi_{i}(t)$ defined in \eqref{eq:dynamic-function-autonomous}.
Then the multi-agent system \eqref{eq:overall-event-trigger} admits
global bipartite consensus. }
\end{thm}
\begin{IEEEproof}
\textcolor{black}{Consider the Lyapunov function candidate as follows,
\begin{equation}
V(t)=V_{1}(t)+V_{2}(t),\label{eq:Lyapunov function for autonomous}
\end{equation}
where 
\[
V_{1}(t)=\boldsymbol{x}^{T}(t)L\boldsymbol{x}(t),
\]
 and 
\[
V_{2}(t)=\sum_{i=1}^{n}\psi_{i}(t).
\]
For any $t\geq0$, from the equations in \eqref{eq:event-triggered-time}
and \eqref{eq:dynamic-function-autonomous}, one has,
\[
\dot{\psi}_{i}(t)\geq-\beta_{i}\psi_{i}(t)-\frac{\delta_{i}}{\theta_{i}}\psi_{i}(t),
\]
and
\[
\psi_{i}(t)\geq\psi_{i}(0)e^{-(\beta_{i}+\frac{\delta_{i}}{\theta_{i}})t}>0,
\]
therefore, one can get that $V(t)\geq0$.}

\textcolor{black}{Computing the time derivative of $V_{1}(t)$ along
with \eqref{eq:overall-event-trigger} yields,
\begin{eqnarray*}
\dot{V}_{1}(t) & = & \dot{\boldsymbol{x}}^{T}(t)L\boldsymbol{x}(t)+\boldsymbol{x}(t)^{T}L\dot{\boldsymbol{x}}(t)\\
 & = & 2\boldsymbol{x}(t)^{T}L\text{{\bf sat}}_{\varDelta}\left(\widehat{\boldsymbol{u}}(t)\right)\\
 & = & -2\boldsymbol{u}(t)^{T}\text{{\bf sat}}_{\varDelta}\left(\widehat{\boldsymbol{u}}(t)\right).
\end{eqnarray*}
Let $\boldsymbol{\phi}(t)=[\boldsymbol{\phi}_{1}^{T}(t),\boldsymbol{\phi}_{2}^{T}(t),\ldots,\boldsymbol{\phi}_{n}^{T}(t)]^{T}\in\mathbb{R}^{dn}$
and $\boldsymbol{\phi}(t)=\widehat{\boldsymbol{u}}(t)-\boldsymbol{u}(t)$,
then one has,
\begin{eqnarray*}
\dot{V}_{1}(t) & = & -2\widehat{\boldsymbol{u}}^{T}(t)\text{{\bf sat}}_{\varDelta}\left(\widehat{\boldsymbol{u}}(t)\right)+2\boldsymbol{\phi}(t)^{T}\text{{\bf sat}}_{\varDelta}\left(\widehat{\boldsymbol{u}}(t)\right)\\
 & = & -\sum_{i=1}^{n}2\widehat{\boldsymbol{u}}_{i}^{T}(t)\text{{\bf sat}}_{\varDelta}\left(\widehat{\boldsymbol{u}}_{i}(t)\right)+\sum_{i=1}^{n}2\boldsymbol{\phi}_{i}(t)^{T}\text{{\bf sat}}_{\varDelta}\left(\widehat{\boldsymbol{u}}_{i}(t)\right)\\
 & \leq & -\sum_{i=1}^{n}2\widehat{\boldsymbol{u}}_{i}^{T}(t)\text{{\bf sat}}_{\varDelta}\left(\widehat{\boldsymbol{u}}_{i}(t)\right)+\sum_{i=1}^{n}\boldsymbol{\phi}_{i}(t)^{T}\boldsymbol{\phi}_{i}(t)\\
 & + & \sum_{i=1}^{n}\text{{\bf sat}}_{\varDelta}\left(\widehat{\boldsymbol{u}}_{i}(t)\right)^{T}\text{{\bf sat}}_{\varDelta}\left(\widehat{\boldsymbol{u}}_{i}(t)\right),
\end{eqnarray*}
according to Lemma \ref{saturation-inequality}, one has,
\[
\text{{\bf sat}}_{\varDelta}\left(\widehat{\boldsymbol{u}}_{i}(t)\right)^{T}\text{{\bf sat}}_{\varDelta}\left(\widehat{\boldsymbol{u}}_{i}(t)\right)\leq\widehat{\boldsymbol{u}}_{i}^{T}(t)\text{{\bf sat}}_{\varDelta}\left(\widehat{\boldsymbol{u}}_{i}(t)\right),
\]
therefore,
\begin{align*}
\dot{V}_{1}(t) & \leq-\sum_{i=1}^{n}\widehat{\boldsymbol{u}}_{i}^{T}(t)\text{{\bf sat}}_{\varDelta}\left(\widehat{\boldsymbol{u}}_{i}(t)\right)+\sum_{i=1}^{n}\boldsymbol{\phi}_{i}(t)^{T}\boldsymbol{\phi}_{i}(t).
\end{align*}
Since,
\begin{eqnarray*}
\boldsymbol{\phi}_{i}(t) & = & \sum_{j\in\mathcal{N}_{i}}|A_{ij}|\left(\text{{\bf sgn}}(A_{ij})\widehat{\boldsymbol{x}}_{j}(t)-\widehat{\boldsymbol{x}}_{i}(t)\right)\\
 &  & -\sum_{j\in\mathcal{N}_{i}}|A_{ij}|(\text{{\bf sgn}}(A_{ij})\boldsymbol{x}_{j}(t)-\boldsymbol{x}_{i}(t))\\
 & = & \sum_{j\in\mathcal{N}_{i}}|A_{ij}|(\text{{\bf sgn}}(A_{ij})\boldsymbol{e}_{j}(t)-\boldsymbol{e}_{i}(t)),
\end{eqnarray*}
thus,}

\textcolor{black}{
\begin{eqnarray*}
\parallel\boldsymbol{\phi}_{i}(t)\parallel & \leq & \sum_{j\in\mathcal{N}_{i}}\parallel A_{ij}\parallel\parallel\boldsymbol{e}_{i}(t)\parallel+\sum_{j\in\mathcal{N}_{i}}\parallel A_{ij}\parallel\parallel\boldsymbol{e}_{j}(t)\parallel\\
 & = & \left(\sum_{j\in\mathcal{N}_{i}}\lambda_{d}(\mid A_{ij}\mid)\right)\parallel\boldsymbol{e}_{i}(t)\parallel+\\
 &  & \sum_{j\in\mathcal{N}_{i}}\lambda_{d}(\mid A_{ij}\mid)\parallel\boldsymbol{e}_{j}(t)\parallel,
\end{eqnarray*}
and}

\textcolor{black}{
\begin{eqnarray*}
\parallel\boldsymbol{\phi}_{i}(t)\parallel^{2} & \leq & \left(\mid\mathcal{N}_{i}\mid+1\right)\left(\sum_{j\in\mathcal{N}_{i}}\lambda_{d}(\mid A_{ij}\mid)\right)^{2}\parallel\boldsymbol{e}_{i}(t)\parallel^{2}\\
 & + & \left(\mid\mathcal{N}_{i}\mid+1\right)\sum_{j\in\mathcal{N}_{i}}\lambda_{d}^{2}(\mid A_{ij}\mid)\parallel\boldsymbol{e}_{j}(t)\parallel^{2}.
\end{eqnarray*}
Hence,
\begin{eqnarray*}
\sum_{i=1}^{n}\boldsymbol{\phi}_{i}(t)^{T}\boldsymbol{\phi}_{i}(t) & \leq & \sum_{i=1}^{n}n\left(\sum_{j\in\mathcal{N}_{i}}\lambda_{d}(\mid A_{ij}\mid)\right)^{2}\parallel\boldsymbol{e}_{i}(t)\parallel^{2}\\
 & + & \sum_{i=1}^{n}n\sum_{j\in\mathcal{N}_{i}}\lambda_{d}^{2}(\mid A_{ij}\mid)\parallel\boldsymbol{e}_{j}(t)\parallel^{2}\\
 & = & \sum_{i=1}^{n}\varpi_{i}\parallel\boldsymbol{e}_{i}(t)\parallel^{2}.
\end{eqnarray*}
where}

\textcolor{black}{
\begin{eqnarray*}
\varpi_{i} & = & n\left(\sum_{j\in\mathcal{N}_{i}}\lambda_{d}(\mid A_{ij}\mid)\right)^{2}+n\sum_{j\in\mathcal{N}_{i}}\lambda_{d}^{2}(\mid A_{ij}\mid).
\end{eqnarray*}
Thus, 
\begin{eqnarray*}
\dot{V}_{1}(t) & \leq & \sum_{i=1}^{n}\varpi_{i}\parallel\boldsymbol{e}_{i}(t)\parallel^{2}-\sum_{i=1}^{n}\widehat{\boldsymbol{u}}_{i}^{T}(t)\text{{\bf sat}}_{\varDelta}\left(\widehat{\boldsymbol{u}}_{i}(t)\right).
\end{eqnarray*}
Now, we are in position to consider the Lyapunov function candidate
$V(t)$ in \eqref{eq:Lyapunov function for autonomous}, one has}

\textcolor{black}{
\begin{eqnarray*}
 &  & \dot{V}(t)=\dot{V_{1}}(t)+\sum_{i=1}^{n}\dot{\psi}_{i}(t)\\
 & \leq & \sum_{i=1}^{n}\varpi_{i}\parallel\boldsymbol{e}_{i}(t)\parallel^{2}-\sum_{i=1}^{n}\widehat{\boldsymbol{u}}_{i}^{T}(t)\text{{\bf sat}}_{\varDelta}\left(\widehat{\boldsymbol{u}}_{i}(t)\right)\\
 & + & \sum_{i=1}^{n}\left(\delta_{i}(\rho_{i}\widehat{\boldsymbol{u}}_{i}^{T}(t)\text{{\bf sat}}_{\varDelta}\left(\widehat{\boldsymbol{u}}_{i}(t)\right)-\varpi_{i}\parallel\boldsymbol{e}_{i}(t)\parallel^{2})\right)\\
 & + & \sum_{i=1}^{n}\left(-\beta_{i}\psi_{i}(t)\right)\\
 & = & -\sum_{i=1}^{n}(1-\delta_{i}\rho_{i})\widehat{\boldsymbol{u}}_{i}^{T}(t)\text{{\bf sat}}_{\varDelta}\left(\widehat{\boldsymbol{u}}_{i}(t)\right)\\
 & + & \sum_{i=1}^{n}(1-\delta_{i})\varpi_{i}\parallel\boldsymbol{e}_{i}(t)\parallel^{2}-\sum_{i=1}^{n}\beta_{i}\psi_{i}(t)\\
 & = & -\sum_{i=1}^{n}\beta_{i}\psi_{i}(t)+\sum_{i=1}^{n}(1-\delta_{i})\varpi_{i}\parallel\boldsymbol{e}_{i}(t)\parallel^{2}\\
 & - & \sum_{i=1}^{n}\widehat{\boldsymbol{u}}_{i}^{T}(t)\text{{\bf sat}}_{\varDelta}\left(\widehat{\boldsymbol{u}}_{i}(t)\right)+\sum_{i=1}^{n}\rho_{i}\widehat{\boldsymbol{u}}_{i}^{T}(t)\text{{\bf sat}}_{\varDelta}\left(\widehat{\boldsymbol{u}}_{i}(t)\right)\\
 & - & \sum_{i=1}^{n}(1-\delta_{i})\rho_{i}\widehat{\boldsymbol{u}}_{i}^{T}(t)\text{{\bf sat}}_{\varDelta}\left(\widehat{\boldsymbol{u}}_{i}(t)\right)\\
 & \leq & -\sum_{i=1}^{n}\left(\beta_{i}-\frac{1-\delta_{i}}{\theta_{i}}\right)\psi_{i}(t)-\\
 &  & \sum_{i=1}^{n}(1-\rho_{i})\widehat{\boldsymbol{u}}_{i}^{T}(t)\text{{\bf sat}}_{\varDelta}\left(\widehat{\boldsymbol{u}}_{i}(t)\right)\\
 & \leq & -\sum_{i=1}^{n}\left(\beta_{i}-\frac{1-\delta_{i}}{\theta_{i}}\right)\psi_{i}(t)\\
 & - & (1-{\displaystyle \max_{i\in\underline{n}}}\left\{ \rho_{i}\right\} )\sum_{i=1}^{n}\widehat{\boldsymbol{u}}_{i}^{T}(t)\text{{\bf sat}}_{\varDelta}\left(\widehat{\boldsymbol{u}}_{i}(t)\right),
\end{eqnarray*}
therefore, $\dot{V}(t)\leq0$. Due to $V(t)\geq0$ and $\dot{V}(t)\leq0$,
which implies that $\underset{t\rightarrow\infty}{\text{{\bf lim}}}\dot{V}(t)=0$.Thus,
one has $\underset{t\rightarrow\infty}{\text{{\bf lim}}}\psi_{i}(t)=0$
and $\underset{t\rightarrow\infty}{\text{{\bf lim}}}\widehat{\boldsymbol{u}}_{i}(t)=\boldsymbol{0}$.
Due to,
\begin{align*}
0 & \leq\parallel\boldsymbol{e}_{i}(t)\parallel^{2}\\
 & \leq\frac{\rho_{i}}{\varpi_{i}}\widehat{\boldsymbol{u}}_{i}^{T}(t)\text{{\bf sat}}_{\varDelta}\left(\widehat{\boldsymbol{u}}_{i}(t)\right)+\frac{1}{\varpi_{i}\theta_{i}}\psi_{i}(t),
\end{align*}
therefore, $\underset{t\rightarrow\infty}{\text{{\bf lim}}}\boldsymbol{e}_{i}(t)=\boldsymbol{0}$.
Then, one has,
\begin{eqnarray*}
\dot{V}_{1}(t) & = & \dot{\boldsymbol{x}}^{T}(t)L\boldsymbol{x}(t)+\boldsymbol{x}(t)^{T}L\dot{\boldsymbol{x}}(t)\\
 & = & -2\boldsymbol{x}(t)^{T}L\text{{\bf sat}}_{\varDelta}\left(L(\boldsymbol{x}(t)+\boldsymbol{e}(t))\right),
\end{eqnarray*}
thus, $\underset{t\rightarrow\infty}{\text{{\bf lim}}}L\boldsymbol{x}(t)=\boldsymbol{0}$
and 
\[
\underset{t\rightarrow\infty}{\text{{\bf lim}}}\left(\boldsymbol{x}_{i}(t)-\text{{\bf sgn}}(A_{ij})\boldsymbol{x}_{j}(t)\right)=\boldsymbol{0},\,\forall i,j\in\underline{n}.
\]
That is, the multi-agent system \eqref{eq:overall-event-trigger}
achieves global bipartite consensus.}

\textcolor{black}{  }
\end{IEEEproof}
\textcolor{black}{}
\begin{rem}
Notably, owing to the nonlinearity induced by actuator saturation,
the multi-agent system does not always achieve average bipartite consensus.
The final consensus value of the network is eventually influenced
by the saturation level $\Delta$. Specifically, the consensus value
is the average (after a proper gauge transformation) of the agents'
states at the last time instance $T_{\text{sf}}$ that there exists
saturated control inputs in the multi-agent system, that is, $D^{*}(\boldsymbol{1}_{n}\otimes(\frac{1}{n}(\boldsymbol{1}_{n}^{T}\otimes I_{d})D^{*}\boldsymbol{x}(T_{\text{sf}})))$.
After $T_{\text{sf}}$, the saturation constraint on multi-agent system
is eliminated until the achievement of final bipartite consensus. 
\end{rem}
\begin{rem}
\textcolor{black}{The proposed event-triggered algorithm for the multi-agent
system with saturation here is not only applicable to the matrix-weighted
networks but also to the scalar-weighted networks. Note that \eqref{eq:the agent protocol}
degenerates into the scalar-weighted case when $A_{ij}=a_{ij}I$ where
$a_{ij}\in\mathbb{R}$ and $I$ denotes the $d\times d$ identity
matrix and in this case, one can choose
\[
\lambda_{d}(\mid A_{ij}\mid)=|a_{ij}|.
\]
Then the triggering function \eqref{eq:event-triggered-time} is also
suitable for the scalar-weighted networks.}
\end{rem}
\textcolor{black}{}
\textcolor{black}{In the following discussion, we shall prove that
Zeno behavior can be avoided using  the aforementioned event-triggered
strategy. We have the following result.}
\begin{thm}
\textcolor{black}{\label{zeno-avoided}Under the global bipartite
consensus condition in the Theorem \ref{event-triggered-leaderless},
the Zeno behavior of multi-agent system \eqref{eq:overall-event-trigger}
under the matrix-weighted network $\mathcal{G}=(\mathcal{V},\mathcal{E},A)$
can be avoided, i.e., there are no infinite triggering instants in
a finite time.}
\end{thm}
\begin{IEEEproof}
\textcolor{black}{By contradiction, suppose that there exists Zeno
behavior. Then, there at least exists one agent $i$ such that $\text{{\bf lim}}_{k\rightarrow\infty}t_{k}^{i}=T_{0}$
where $T_{0}>0$. From the above analysis, we know that there exists
a positive constant $M_{0}>0$ satisfying $\parallel\boldsymbol{x}_{i}(t)\parallel\leq M_{0}$
for all $t\geq0$ and $i\in\underline{n}$. Then one has 
\[
\parallel\boldsymbol{u}_{i}(t)\parallel\leq2M_{o}\sum_{j\in\mathcal{N}_{i}}\lambda_{d}(\mid A_{ij}\mid),
\]
 for any $t\geq0$. Choose 
\begin{align*}
\varepsilon_{0} & =(4M_{o}\sum_{j\in\mathcal{N}_{i}}\lambda_{d}(\mid A_{ij}\mid))^{-1}\sqrt{\frac{\psi_{i}(0)}{\theta_{i}\varpi_{i}}}e^{-\frac{1}{2}(\beta_{i}+\frac{\delta_{i}}{\theta_{i}})T_{0}},
\end{align*}
according to the definition of limits, there exists a positive integer
$N(\varepsilon_{0})$ such that for any $k\geq N(\varepsilon_{0})$,
\begin{equation}
t_{k}^{i}\in[T_{0}-\varepsilon_{0},T_{0}].\label{eq:the limit}
\end{equation}
Then one sufficient condition to guarantee the inequality in \eqref{eq:event-triggered-time}
is}

\textcolor{black}{
\[
\parallel\boldsymbol{e}_{i}(t)\parallel\leq\sqrt{\frac{\psi_{i}(0)}{\theta_{i}\varpi_{i}}}e^{-\frac{1}{2}(\beta_{i}+\frac{\delta_{i}}{\theta_{i}})t}.
\]
In addition, 
\begin{eqnarray*}
\parallel\boldsymbol{e}_{i}(t)\parallel & = & \parallel\widehat{\boldsymbol{x}}_{i}(t_{k}^{i})-\boldsymbol{x}_{i}(t)\parallel\\
 & = & \parallel\boldsymbol{x}_{i}(t_{k}^{i})-\boldsymbol{x}_{i}(t)\parallel\\
 & = & \left\Vert \int_{t_{k}^{i}}^{t}\dot{\boldsymbol{x}}_{i}(t)d(t)\right\Vert \\
 & \leq & \int_{t_{k}^{i}}^{t}\parallel\dot{\boldsymbol{x}}_{i}(t)\parallel d(t)\\
 & \leq & (t-t_{k}^{i})\left(2M_{o}\sum_{j\in\mathcal{N}_{i}}\lambda_{d}(\mid A_{ij}\mid)\right),
\end{eqnarray*}
then another sufficient condition to guarantee that the inequality
in \eqref{eq:event-triggered-time} holds if
\begin{align}
 & (t-t_{k}^{i})\left(2M_{o}\sum_{j\in\mathcal{N}_{i}}\lambda_{d}(\mid A_{ij}\mid)\right)\nonumber \\
\leq & \sqrt{\frac{\psi_{i}(0)}{\theta_{i}\varpi_{i}}}e^{-\frac{1}{2}(\beta_{i}+\frac{\delta_{i}}{\theta_{i}})t},\label{eq:zeno-event-triggered-time}
\end{align}
Let $t_{N(\varepsilon_{0})+1}^{i}$ and $\tilde{t}_{N(\varepsilon_{0})+1}^{i}$
denote the next triggering time determined by the inequalities in
\eqref{eq:event-triggered-time} and \eqref{eq:zeno-event-triggered-time},
respectively. Then, }

\textcolor{black}{
\begin{eqnarray*}
 &  & t_{N(\varepsilon_{0})+1}^{i}-t_{N(\varepsilon_{0})}^{i}\\
 & \geq & \tilde{t}_{N(\varepsilon_{0})+1}^{i}-t_{N(\varepsilon_{0})}^{i}\\
 & = & \left(2M_{o}\sum_{j\in\mathcal{N}_{i}}\lambda_{d}(\mid A_{ij}\mid)\right)^{-1}\sqrt{\frac{\psi_{i}(0)}{\theta_{i}\varpi_{i}}}e^{-\frac{1}{2}(\beta_{i}+\frac{\delta_{i}}{\theta_{i}})\tilde{t}_{N(\varepsilon_{0})+1}^{i}}\\
 & \geq & \left(2M_{o}\sum_{j\in\mathcal{N}_{i}}\lambda_{d}(\mid A_{ij}\mid)\right)^{-1}\sqrt{\frac{\psi_{i}(0)}{\theta_{i}\varpi_{i}}}e^{-\frac{1}{2}(\beta_{i}+\frac{\delta_{i}}{\theta_{i}})T_{0}}\\
 & = & 2\varepsilon_{0},
\end{eqnarray*}
which contradicts with the equation in \eqref{eq:the limit}. Therefore,
Zeno behavior is excluded.}
\end{IEEEproof}
\textcolor{black}{}

\section{\textcolor{black}{Leader-follower Matrix-weighted Networks}}

\textcolor{black}{Besides the leaderless network, there also exists
another popular paradigm where a subset of agents are selected as
leaders or informed agents to steer the network state to a desired
one which is referred to as leader-follower network. In a leader-follower
network, a subset of agents are referred to as leaders (or informed
agents), denoted by $\mathcal{V}_{\text{leader}}\subset\mathcal{V}$,
who can be directly influenced by the external input signal, the remaining
agents are referred to as followers, denoted by $\mathcal{V}_{\text{follower}}=\mathcal{V}\setminus\mathcal{V}_{\text{leader}}$.
The set of external input signal is denoted by $\mathcal{W}=\left\{ \boldsymbol{w}_{1},\dots,\boldsymbol{w}_{m}\right\} $
where $\boldsymbol{w}_{l}\in\mathbb{R}^{d}$, $l\in\underline{m}$
and $m\in\mathbb{Z}_{+}$. In the following discussion, we shall assume
that the input signal is homogeneous, i.e., $\boldsymbol{w}_{l_{1}}=\boldsymbol{w}_{l_{2}}=\boldsymbol{w}_{0}\ \text{for all \ensuremath{l_{1},l_{2}\in\underline{m}}}$.
Denote by the edge set between external input signals and the leaders
as $\mathcal{E}^{'}$, and a corresponding set of matrix weights as
$B=[B_{il}]\in\mathbb{R}^{nd\times md}$ where $|B_{il}|\geq0$ or
$|B_{il}|>0$ if agent $i$ is influenced by the input $\boldsymbol{w}_{l}$
and $B_{il}=0_{d\times d}$ otherwise. The graph $\bar{\mathcal{G}}=(\bar{\mathcal{V}},\bar{\mathcal{E}},\bar{A})$
is directed with $\bar{\mathcal{V}}=\mathcal{V}\cup\mathcal{W}$,
$\bar{\mathcal{E}}=\mathcal{E}\cup\mathcal{E}^{'}$, $\bar{A}=A\cup B$. }

\subsection{\textcolor{black}{Actuator Saturation}}

Similar to the leaderless case, we now first consider the following
leader-follower control protocol without the event-triggered communication
constraint,\textcolor{black}{
\begin{equation}
\dot{\boldsymbol{x}}_{i}(t)=\text{{\bf sat}}_{\varDelta}(\boldsymbol{q}_{i}(t)),i\in\mathcal{V},\label{eq:the agent protocol-1}
\end{equation}
where
\begin{align}
\boldsymbol{q}_{i}(t) & =-\sum_{j\in\mathcal{N}_{i}}|A_{ij}|(\boldsymbol{x}_{i}(t)-\text{{\bf sgn}}(A_{ij})\boldsymbol{x}_{j}(t))\nonumber \\
 & -\sum_{l=1}^{m}|B_{il}|(\boldsymbol{x}_{i}(t)-\text{{\bf sgn}}(B_{il})\boldsymbol{w}_{l}),i\in\mathcal{V}\text{.}\label{eq:LF-protocol}
\end{align}
The collective dynamics of \eqref{eq:the agent protocol-1} can subsequently
be characterized by 
\begin{equation}
\dot{\boldsymbol{x}}=\text{{\bf sat}}_{\varDelta}(-L_{B}(\mathcal{G})\boldsymbol{x}+B\boldsymbol{w}),\label{eq: signed-LF-overall}
\end{equation}
where $\boldsymbol{x}=(\boldsymbol{x}_{1}^{T}(t),\boldsymbol{x}_{2}^{T}(t),\dots,\boldsymbol{x}_{n}^{T}(t))^{T}\in\mathbb{R}^{nd}$,
$\boldsymbol{w}=(\boldsymbol{w}_{1}^{T},\boldsymbol{w}_{2}^{T},\dots,\boldsymbol{w}_{m}^{T})^{T}\in\mathbb{R}^{md}$
and 
\[
L_{B}(\mathcal{G})=L(\mathcal{G})+\text{{\bf blkdiag}}(\sum_{l=1}^{m}|B_{il}|).
\]
}
\begin{defn}
\textcolor{black}{For $i\in\mathcal{V}$ and an arbitrary $\boldsymbol{x_{i}}(0){\color{red}{\color{black}\in\mathbb{R}^{d}}}$,
the multi-agent system \eqref{eq: signed-LF-overall} is said to admit
global bipartite leader-follower consensus if ${\color{black}{\color{blue}{\color{black}\lim{}_{t\rightarrow\infty}\mid\boldsymbol{x}_{i}(t)\mid=\mid\boldsymbol{w}_{0}\mid}}}$.}
\end{defn}
\textbf{\textcolor{black}{Assumption}}\textcolor{black}{{} 2.\label{null-space-SB-1}
The matrix-weighted network $\bar{\mathcal{G}}=(\bar{\mathcal{V}},\bar{\mathcal{E}},\bar{A})$
is structurally balanced and ${\displaystyle \sum_{i=1}^{n}}{\displaystyle \sum_{l=1}^{m}}|B_{il}|$
is positive definite.}
\begin{rem}
\textcolor{black}{The Assumption 1 and Assumption 2 together guarantee
that the leader-follower multi-agent network \eqref{eq: signed-LF-overall}
without saturation admits a bipartite leader-follower consensus \cite{pan2018bipartite,trinh2017matrix}. }
\end{rem}
\textcolor{black}{}
\textcolor{black}{In the following, we shall analyze the convergence
situation of the leader-follower multi-agent system \eqref{eq: signed-LF-overall}
on matrix-weighted networks.}
\begin{lem}
\textcolor{black}{Let Assumptions 1 and 2 hold. Then, the multi-agent
system \eqref{eq: signed-LF-overall} achieves global bipartite leader-follower
consensus.}
\end{lem}
\begin{IEEEproof}
\textcolor{black}{Let 
\[
\boldsymbol{\xi}(t)=\boldsymbol{x}(t)-D^{*}(\boldsymbol{1}_{n}\otimes\boldsymbol{w}_{0}),
\]
where $D^{*}$ is the gauge transformation corresponding to the matrix-weighted
network. Then one has,}

\textcolor{black}{
\begin{equation}
\dot{\boldsymbol{\xi}}(t)=\text{{\bf sat}}_{\varDelta}(-L_{B}\boldsymbol{\xi}(t)).\label{eq:LF-error-system}
\end{equation}
Consider the Lyapunov function candidate as follows,
\[
V(t)=\frac{1}{2}\boldsymbol{\xi}^{T}(t)L_{B}\boldsymbol{\xi}(t),
\]
computing the time derivative of $V(t)$ along with \eqref{eq:LF-error-system}
yields,
\begin{eqnarray*}
\dot{V}(t) & = & \boldsymbol{\xi}(t)^{T}L_{B}\dot{\boldsymbol{\xi}}(t)\\
 & = & \boldsymbol{\xi}(t)^{T}L_{B}\text{{\bf sat}}_{\varDelta}(-L_{B}\boldsymbol{\xi}(t))\leq0.
\end{eqnarray*}
It is obvious that $\dot{V}(t)=0$ if and only if $L_{B}\boldsymbol{\xi}(t)=\boldsymbol{0}$,
i.e., $\boldsymbol{\xi}(t)=\boldsymbol{0}$. Thus according to LaSalle's
invariance principle, 
\[
\underset{t\rightarrow\infty}{\text{{\bf lim}}}\left(\mid\boldsymbol{x}_{i}(t)\mid-\mid\boldsymbol{w}_{0}\mid\right)=\boldsymbol{0},\,\forall i\in\underline{n}.
\]
That is, the multi-agent system \eqref{eq: signed-LF-overall} achieves
global bipartite leader-follower consensus.}
\end{IEEEproof}

\subsection{\textcolor{black}{Event-triggered Mechanism Design}}

\textcolor{black}{In order to avoid continuously information exchange
amongst agents and updating actuators, we proceed to equip the protocol
\eqref{eq: signed-LF-overall} with an event-triggered communication
mechanism. Consider the following protocol for leader-follower multi-agent
system with input saturation and event-triggered constraint,}

\textcolor{black}{
\begin{equation}
\dot{\boldsymbol{x}}_{i}(t)=\text{{\bf sat}}_{\varDelta}(\widehat{\boldsymbol{q}}_{i}(t)),i\in\mathcal{V},\label{eq:leader-follower-event-triggered-saturation}
\end{equation}
where
\begin{align}
\widehat{\boldsymbol{q}}_{i}(t) & =-\sum_{j\in\mathcal{N}_{i}}|A_{ij}|(\widehat{\boldsymbol{x}}_{i}(t)-\text{{\bf sgn}}(A_{ij})\widehat{\boldsymbol{x}}_{j}(t))\nonumber \\
 & -\sum_{l=1}^{m}|B_{il}|(\widehat{\boldsymbol{x}}_{i}(t)-\text{{\bf sgn}}(B_{il})\boldsymbol{w}_{l}),i\in\mathcal{V}.\label{eq:leader-follower-event-triggered-protocol}
\end{align}
The collective dynamics of \eqref{eq:leader-follower-event-triggered-saturation}
can subsequently be characterized by,
\begin{equation}
\dot{\boldsymbol{x}}(t)=\text{{\bf sat}}_{\varDelta}(-L_{B}\widehat{\boldsymbol{x}}(t)+B\boldsymbol{w}),\label{eq:LF-overall-event-trigger}
\end{equation}
where $\widehat{\boldsymbol{x}}(t)=[\widehat{\boldsymbol{x}}_{1}^{T}(t),\widehat{\boldsymbol{x}}_{2}^{T}(t),\ldots,\widehat{\boldsymbol{x}}_{n}^{T}(t)]^{T}\in\mathbb{R}^{dn}.$
Define the state-based measurement error between the last broadcast
state of agent $i\in\mathcal{V}$ and its current state at time $t\geq0$
as
\[
\boldsymbol{e}_{i}(t)=\widehat{\boldsymbol{x}}_{i}(t)-\boldsymbol{x}_{i}(t),
\]
then the system-wise measurement error is denoted by $\boldsymbol{e}(t)=[\boldsymbol{e}_{1}^{T}(t),\boldsymbol{e}_{2}^{T}(t),\ldots,\boldsymbol{e}_{n}^{T}(t)]^{T}$.
For agent $i\in\mathcal{V}$, the triggering time sequence is initiated
from $t_{1}^{i}=0$ and subsequently determined by,}

\textcolor{black}{
\begin{align}
t_{k+1}^{i} & =\underset{r\geq t_{k}^{i}}{\text{{\bf max}}}\{r\thinspace|\thinspace\theta_{i}(\omega_{i}\parallel\boldsymbol{e}_{i}(t)\parallel^{2}-\rho_{i}\widehat{\boldsymbol{q}}_{i}^{T}(t)\text{{\bf sat}}_{\varDelta}\left(\widehat{\boldsymbol{q}}_{i}(t)\right))\nonumber \\
 & \leq\psi_{i}(t),\,\forall t\in[t_{k}^{i},r]\},\nonumber \\
\label{eq:event-triggered-time-1}
\end{align}
where $k\in\mathbb{Z}_{+}$, $\rho_{i}\in[0,1)$, $\theta_{i}$ and
$\varpi_{i}$ are the design parameters and $\psi_{i}(t)$ is an auxiliary
system for each agent $i\in\mathcal{V}$ such that
\begin{eqnarray}
\dot{\psi}_{i}(t) & = & -\beta_{i}\psi_{i}(t)+\delta_{i}(\rho_{i}\widehat{\boldsymbol{q}}_{i}^{T}(t)\text{{\bf sat}}_{\varDelta}\left(\widehat{\boldsymbol{q}}_{i}(t)\right)\nonumber \\
 &  & -\omega_{i}\parallel\boldsymbol{e}_{i}(t)\parallel^{2}),\label{eq: dynamic-function-1}
\end{eqnarray}
with $\psi_{i}(0)>0$, $\beta_{i}>0$ and $\delta_{i}\in[0,1]$.}
\begin{thm}
\textcolor{black}{\label{thm:event-triggered-theorem-for-leader-follower}Consider
the multi-agent system \eqref{eq:LF-overall-event-trigger} under
the matrix-weighted network $\mathcal{G}=(\mathcal{V},\mathcal{E},A)$
satisfying Assumptions 1 and 2. Let $\theta_{i}$ and $\varpi_{i}$
be such that $\theta_{i}>\frac{1-\delta_{i}}{\beta_{i}}$ and 
\begin{eqnarray*}
\omega_{i} & = & n\left(\sum_{j\in\mathcal{N}_{i}}\lambda_{d}(\mid A_{ij}\mid)+\sum_{l=1}^{m}\lambda_{d}(\mid B_{il}\mid)\right)^{2}\\
 & + & n\sum_{j\in\mathcal{N}_{i}}\lambda_{d}^{2}(\mid A_{ij}\mid).
\end{eqnarray*}
for all $i\in\mathcal{V}$, the triggering time sequence is determined
by \eqref{eq:event-triggered-time-1} for agent $i$ with $\psi_{i}(t)$
defined in \eqref{eq: dynamic-function-1}. Then the multi-agent system
\eqref{eq:LF-overall-event-trigger} admits a global bipartite leader-follower
consensus. Moreover, there is no Zeno behavior.}
\end{thm}
\begin{IEEEproof}
\textcolor{black}{Let $\boldsymbol{\xi}(t)=\boldsymbol{x}(t)-D^{*}(\boldsymbol{1}_{n}\otimes\boldsymbol{w}_{0})$,
where $D^{*}$ is the gauge transformation corresponding to the matrix-weighted
network $\mathcal{G}=(\mathcal{V},\mathcal{E},A)$. Then one has,}

\textcolor{black}{
\begin{equation}
\dot{\boldsymbol{\xi}}(t)=-L_{B}\boldsymbol{\xi}(t).\label{eq:LF-error-system-1-1}
\end{equation}
Consider the Lyapunov function candidate as follows,
\[
V(t)=V_{1}(t)+V_{2}(t),
\]
where 
\[
V_{1}(t)=\boldsymbol{\xi}^{T}(t)L_{B}\boldsymbol{\xi}(t),
\]
 and 
\[
V_{2}(t)=\sum_{i=1}^{n}\psi_{i}(t).
\]
}

\textcolor{black}{Different from the leaderless case, in the leader-follower
situation, denote by $\boldsymbol{\phi}_{i}(t)=\widehat{\boldsymbol{q}}_{i}(t)-\boldsymbol{q}_{i}(t)$
and $\boldsymbol{\phi}(t)=\left(\boldsymbol{\phi}_{1}^{T}(t),\boldsymbol{\phi}_{2}^{T}(t),\dots,\boldsymbol{\phi}_{n}^{T}(t)\right)^{T}$,
where $\boldsymbol{q}_{i}(t)$ and $\widehat{\boldsymbol{q}}_{i}(t)$
are defined in \eqref{eq:LF-protocol} and \eqref{eq:leader-follower-event-triggered-protocol},
respectively. Then one has,
\[
\parallel\boldsymbol{\phi}_{i}(t)\parallel^{2}\leq\omega_{i}\parallel\boldsymbol{e}_{i}(t)\parallel^{2},
\]
where 
\begin{eqnarray*}
\omega_{i} & = & n\left(\sum_{j\in\mathcal{N}_{i}}\lambda_{d}(\mid A_{ij}\mid)+\sum_{l=1}^{m}\lambda_{d}(\mid B_{il}\mid)\right)^{2}\\
 & + & n\sum_{j\in\mathcal{N}_{i}}\lambda_{d}^{2}(\mid A_{ij}\mid).
\end{eqnarray*}
 Then, similar to the proof of Theorem \ref{event-triggered-leaderless}
and Theorem \ref{zeno-avoided}, one can get the conclusion.}
\end{IEEEproof}
\textcolor{black}{}
\begin{figure}[tbh]
\begin{centering}
\begin{tikzpicture}[scale=1]
	\node (n1) at (2.4,1.8) [circle,fill=black!0,draw] {1};
    \node (n2) at (4,2.8) [circle,fill=black!0,draw] {2};
    \node (n3) at (5.5,1.8) [circle,fill=black!0,draw] {3};
	\node (n4) at (5,0) [circle,fill=black!0,draw] {4};
    \node (n5) at (3,0) [circle,fill=black!0,draw] {5};

	\node (G) at (4,-0.8) {$\mathcal{G}_1$};


	\draw [<->, thick, color=blue!60] (n1) -- (n2); 
	\draw [<->,thick, color=blue!60] (n1) -- (n5); 
	\draw [<->,thick, color=blue!60] (n3) -- (n4); 
	\draw [<->,thick, color=blue!60,dashed] (n2) -- (n5); 
	\draw[<->,thick, color=red!60] (n2) -- (n3); 

	\draw [<->,thick, color=red!60,,dashed] (n5) -- (n4); 
\end{tikzpicture}
\par\end{centering}
\caption{A structurally balanced matrix-weighted network $\mathcal{G}_{1}$. }

\label{fig:Figure1-1}
\end{figure}
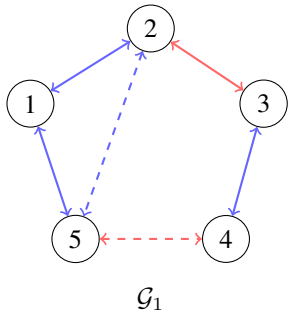

\begin{rem}
\textcolor{black}{Similar to the leaderless case, the event-triggered
strategy proposed for the matrix-weighted leader-follower system with
saturation can be applied for the scalar-weighted leader-follower
case directly. }
\begin{figure}[h]
\begin{centering}
\includegraphics[width=8.5cm]{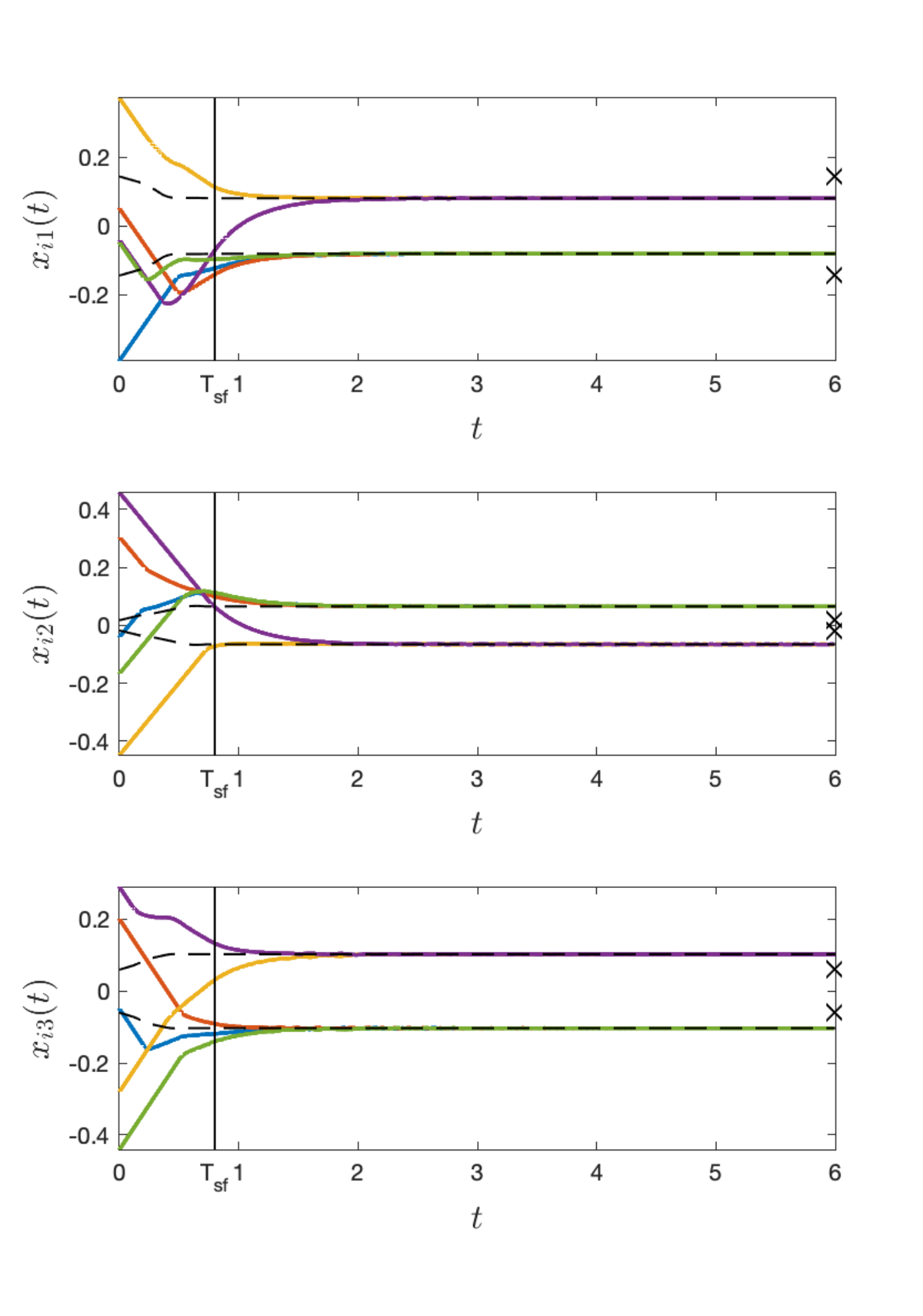}
\par\end{centering}
\caption{Entry-wise trajectory of each agent for the multi-agent system \eqref{eq:overall-event-trigger}
under the structurally balanced matrix-weighted network $\mathcal{G}_{1}$
in Figure \ref{fig:Figure1-1}. }
\label{fig:LL-trajectory}
\end{figure}
 
\end{rem}

\section{\textcolor{black}{Simulations}}

\textcolor{black}{In this section, we proceed to provide simulation
examples to demonstrate the effectiven}ess of the proposed event-triggered
coordination strategy. 

\subsection{Leaderless Matrix-weighted Networks}

First, consider the leaderless multi-agent system \eqref{eq:overall-event-trigger}
on the structurally balanced matrix-weighted network $\mathcal{G}_{1}$
in Figure \ref{fig:Figure1-1}. The solid lines represent the edges
weighted by (positive or negative) definite matrices, the dashed lines
represent the edges weighted by (positive or negative) semi-definite
matrices. The blue lines represent edges weighted by positive (semi-)definite
matrices, and red lines represent edges weighted by negative (semi-)definite
matrices. The node bipartition of $\mathcal{G}_{1}$ is $\mathcal{V}_{1}=\{1,2,5\}$
and $\mathcal{V}_{2}=\{3,4\}$. 

In this example, the state dimension of each agent is $d=3$, and
all agents adopt event-triggered control protocol \eqref{eq:agent-event-triggered-protocol}.
The edges in $\mathcal{G}_{1}$ are weighted by

\textcolor{black}{
\[
A_{12}=\begin{bmatrix}10.14 & 1.64 & -2.16\\
1.641 & \text{10.06} & -1.58\\
-2.16 & -1.58 & 12.45
\end{bmatrix}>0,
\]
}

\textcolor{black}{
\[
A_{23}=\begin{bmatrix}-9.75 & 1.87 & 4.69\\
1.87 & -7.17 & 0.72\\
4.69 & 0.72 & -9.51
\end{bmatrix}<0,
\]
}

\textcolor{black}{
\[
A_{15}=\begin{bmatrix}12.42 & -1.51 & -1.07\\
-1.51 & 11.52 & -1.1\\
-1.07 & -1.1 & 14.4
\end{bmatrix}>0,
\]
}

\textcolor{black}{
\[
A_{25}=\begin{bmatrix}3.03 & -2.21 & 3.92\\
-2.21 & 4.58 & -1.63\\
3.92 & -1.63 & 5.6
\end{bmatrix}\ge0,
\]
\[
A_{34}=\begin{bmatrix}7.36 & 4.67 & 5.13\\
4.67 & 10.89 & -2.31\\
5.13 & -2.31 & 9.92
\end{bmatrix}>0,
\]
and
\[
A_{45}=\begin{bmatrix}-4.88 & -3.07 & 0.46\\
-3.07 & -2.82 & -2.03\\
0.46 & -2.03 & -6.13
\end{bmatrix}\le0.
\]
}
\begin{figure}[H]
\begin{centering}
\includegraphics[width=8.5cm]{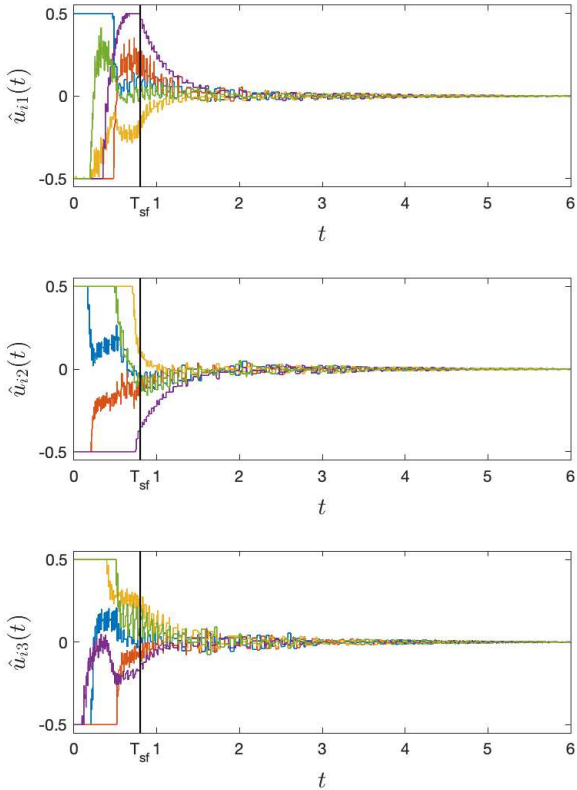}
\par\end{centering}
\caption{The event-based control protocol $\widehat{\boldsymbol{u}}_{i}(t)$
of each agent $i\in\mathcal{V}$ for the multi-agent system \eqref{eq:overall-event-trigger}
under the structurally balanced matrix-weighted network $\mathcal{G}_{1}$. }

\label{fig:LL-trigger-control}
\end{figure}
\textcolor{black}{Moreover, $A_{ij}=A_{ji}$ for all $(i,j)\in\mathcal{E}(\mathcal{G}_{1})$.
Let the saturation level be $\triangle=0.5$. Choose $\rho_{i}=0.9$,
$\delta_{i}=1$, $\beta_{i}=1$, and $\psi_{i}(0)=0.5$. By computing
the eigenvalues of the weight matrices, one has }$\varpi_{1}=6620$,
$\varpi_{2}=10212$, $\varpi_{3}=6355$, $\varpi_{4}=3880$ and $\varpi_{5}=7144$.\textcolor{black}{{}
According to Theorem \ref{event-triggered-leaderless}, one can choose
$\theta_{i}=0.5$ which satisfies $\theta_{i}>\frac{1-\delta_{i}}{\beta_{i}}$.
Each dimension of initial value corresponding to each agent is randomly
chosen from the interval $[-1,1]$. Using the above parameters, the
global bipartite consensus can be achieved in an element-wise manner,
as shown in Figure \ref{fig:LL-trajectory}. The dimensions of control
protocol for each agent are illustrated in Figure \ref{fig:LL-trigger-control}.
 Sequences of triggering time for each agent are illustrated in Figure
\ref{fig:LL-trigger-time}.}

\begin{figure}[H]
\begin{centering}
\includegraphics[width=9cm]{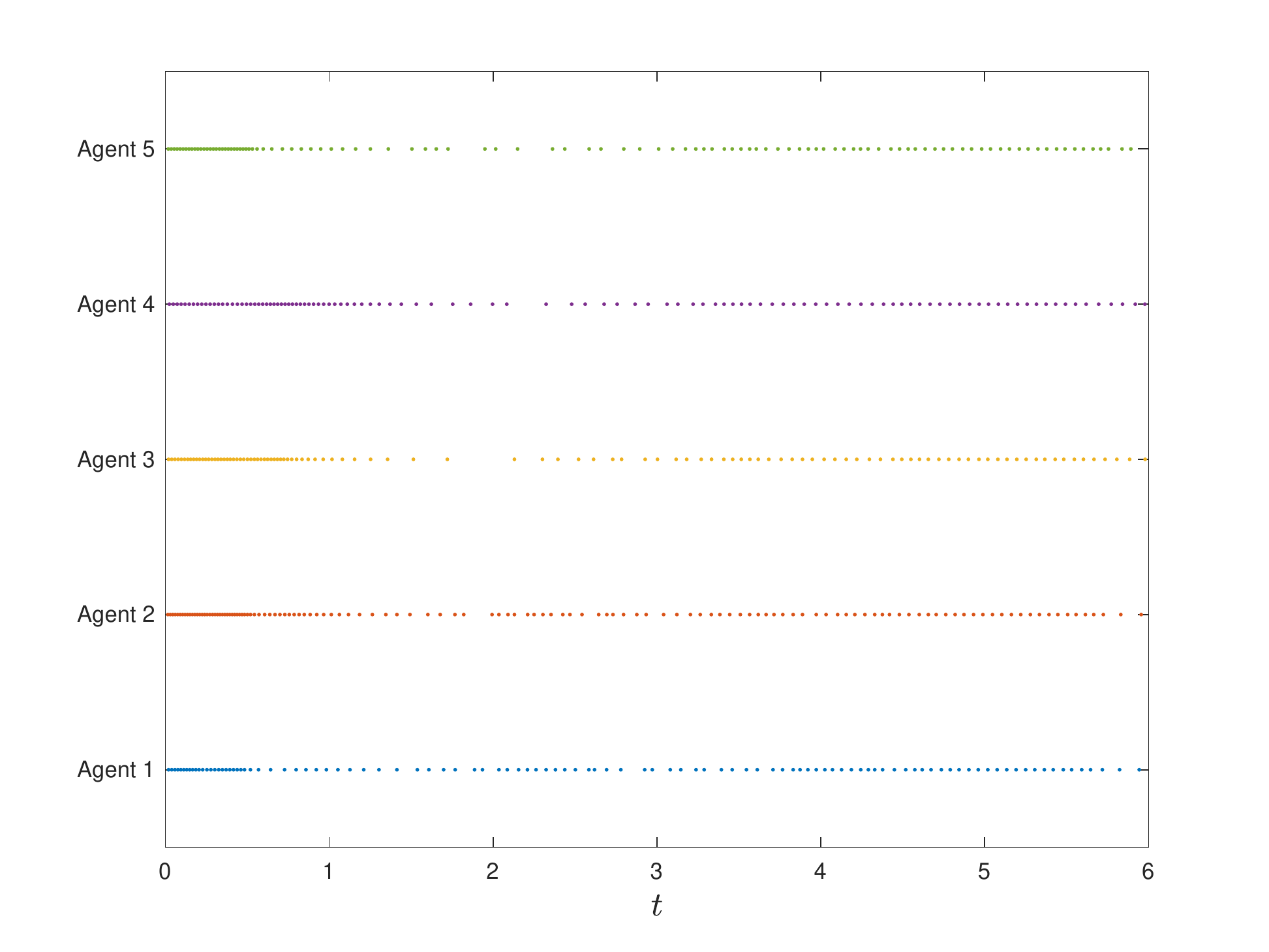}
\par\end{centering}
\caption{The triggering time instants of each agent in the multi-agent system
\eqref{eq:overall-event-trigger} under the structurally balanced
matrix-weighted network $\mathcal{G}_{1}$.}

\label{fig:LL-trigger-time}
\end{figure}

Note that the multi-agent system does not achieve average bipartite
consensus, indicated by black crosses at $t=6$ in Figure \ref{fig:LL-trajectory}.
The simultaneous average (after a proper gauge transformation) of
agents' states is shown in separate dimension in Figure \ref{fig:LL-trajectory},
highlighted by black dashed lines in each panel. Note that this average
value may vary when the each agent is driven by saturated input. As
one can observe that, the multi-agent system behaves in a manner of
saturation-free after $t=T_{\text{sf}}=0.8$. In this example, the
final bipartite consensus value is the average (after a proper gauge
transformation) of $\boldsymbol{x}(0.8)$, namely, $D^{*}(\boldsymbol{1}_{n}\otimes(\frac{1}{n}(\boldsymbol{1}_{n}^{T}\otimes I_{d})D^{*}\boldsymbol{x}(0.8)))$.
The black solid vertical line in each panel of Figure \ref{fig:LL-trajectory}
and Figure \ref{fig:LL-trigger-control} indicates the $T_{\text{sf}}$,
namely, the last time instance that there exists saturated control
inputs in the multi-agent system.

\subsection{\textcolor{black}{Leader-follower Matrix-weighted Networks}}

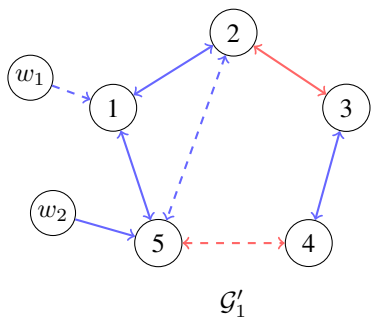
\begin{figure}[h]
\begin{centering}
\begin{tikzpicture}[scale=1]
    \node (n1) at (2.4,1.8) [circle,fill=black!0,draw] {1};
    \node (n2) at (4,2.8) [circle,fill=black!0,draw] {2};
    \node (n3) at (5.5,1.8) [circle,fill=black!0,draw] {3};
	\node (n4) at (5,0) [circle,fill=black!0,draw] {4};
    \node (n5) at (3,0) [circle,fill=black!0,draw] {5};

    \node (u1) at (1.3,2.2) [circle,inner sep= 1.5pt,fill=black!0,draw] {$w_1$};
    \node (u2) at (1.6,0.4) [circle,inner sep= 1.5pt,fill=black!0,draw] {$w_2$};

	\draw[->,thick, dashed, color=blue!60]  (u1) -- (n1); 
	\draw[->,thick, color=blue!60]  (u2) -- (n5);

	\node (G) at (4,-0.8) {$\mathcal{G}^{\prime}_1$};


	\draw [<->,thick, color=blue!60] (n1) -- (n2); 
	\draw [<->,thick, color=blue!60] (n1) -- (n5); 
	\draw [<->,thick, color=blue!60] (n3) -- (n4); 
	\draw [<->,thick, color=blue!60,dashed] (n2) -- (n5); 
	\draw[<->,thick, color=red!60] (n2) -- (n3); 

	\draw [<->,thick, color=red!60,,dashed] (n5) -- (n4);

\end{tikzpicture}
\par\end{centering}
\caption{A structurally balanced matrix-weighted network with two external
inputs $w_{1}$ and $w_{2}$, denoted by $\mathcal{G}_{1}^{\prime}$.
The correspondence between the line pattern and weight matrix is the
same as that in Figure \ref{fig:Figure1-1}.}

\label{fig:LF-network}
\end{figure}

\textcolor{black}{Consider the leader-follower multi-agent system
\eqref{eq:LF-overall-event-trigger} on the leader-follower network
$\mathcal{G}_{1}^{\prime}$ in Figure \ref{fig:LF-network}, where
agents $1$ and $5$ are the leaders influenced by the inputs $\boldsymbol{w}_{1}$
and $\boldsymbol{w}_{2}$, respectively. The edge weights on the matrix-weighted
network $\mathcal{G}_{1}^{\prime}$ are the same as the leaderless
case above, the influence weights by the inputs $\boldsymbol{w}_{1}$
and $\boldsymbol{w}_{2}$ are $B_{11}=A_{25}\ge0,$ and $B_{52}=A_{12}>0,$
respectively.} 
\textcolor{black}{In this case, choose $\rho_{i}=0.9$, $\delta_{i}=1$,
$\beta_{i}=1$, $\psi_{i}(0)=0.5$, and $\boldsymbol{w}_{1}=\boldsymbol{w}_{2}=[0.2,0.4,0.6]^{T}$.
Let the saturation constraint be $\triangle=0.5$. By computing the
eigenvalues of the weight matri}ces, one has $\omega_{1}=10004$,
$\omega_{2}=10212$, $\omega_{3}=6355$, $\omega_{4}=3880$, $\omega_{5}=13027$.
\begin{figure}[h]
\begin{centering}
\includegraphics[width=8.5cm]{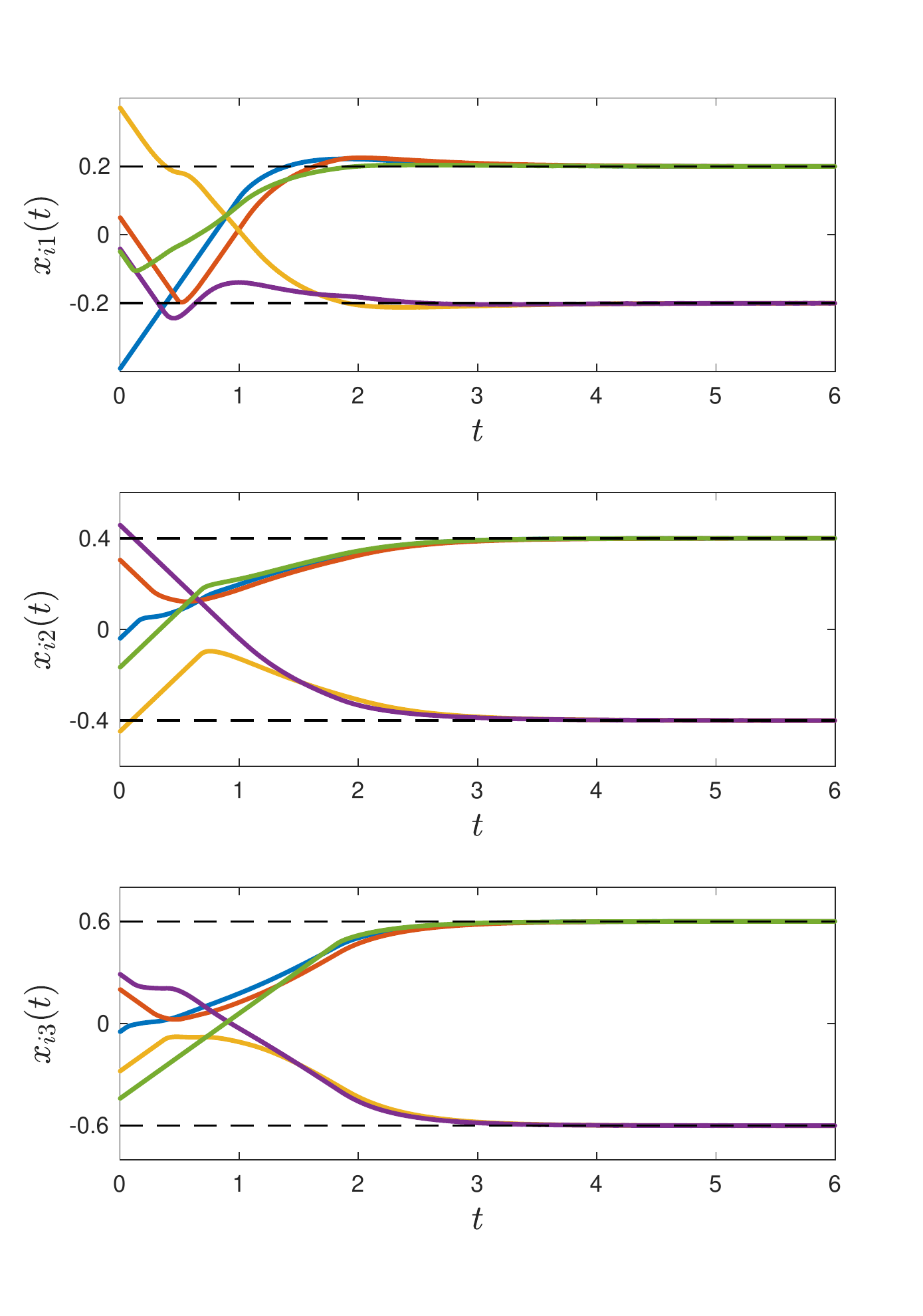}
\par\end{centering}
\caption{Entry-wise trajectory of each agent for the multi-agent system \eqref{eq:LF-overall-event-trigger}
under the leader-follower network $\mathcal{G}_{1}^{\prime}$ in Figure
\ref{fig:LF-network}.}

\label{fig:LF-trajectory}
\end{figure}
\textcolor{black}{According to Theorem \ref{thm:event-triggered-theorem-for-leader-follower},
choose $\theta_{i}=1$ satisfying $\theta_{i}>\frac{1-\delta_{i}}{\beta_{i}}$.
}
\begin{figure}[h]
\begin{centering}
\includegraphics[width=9cm]{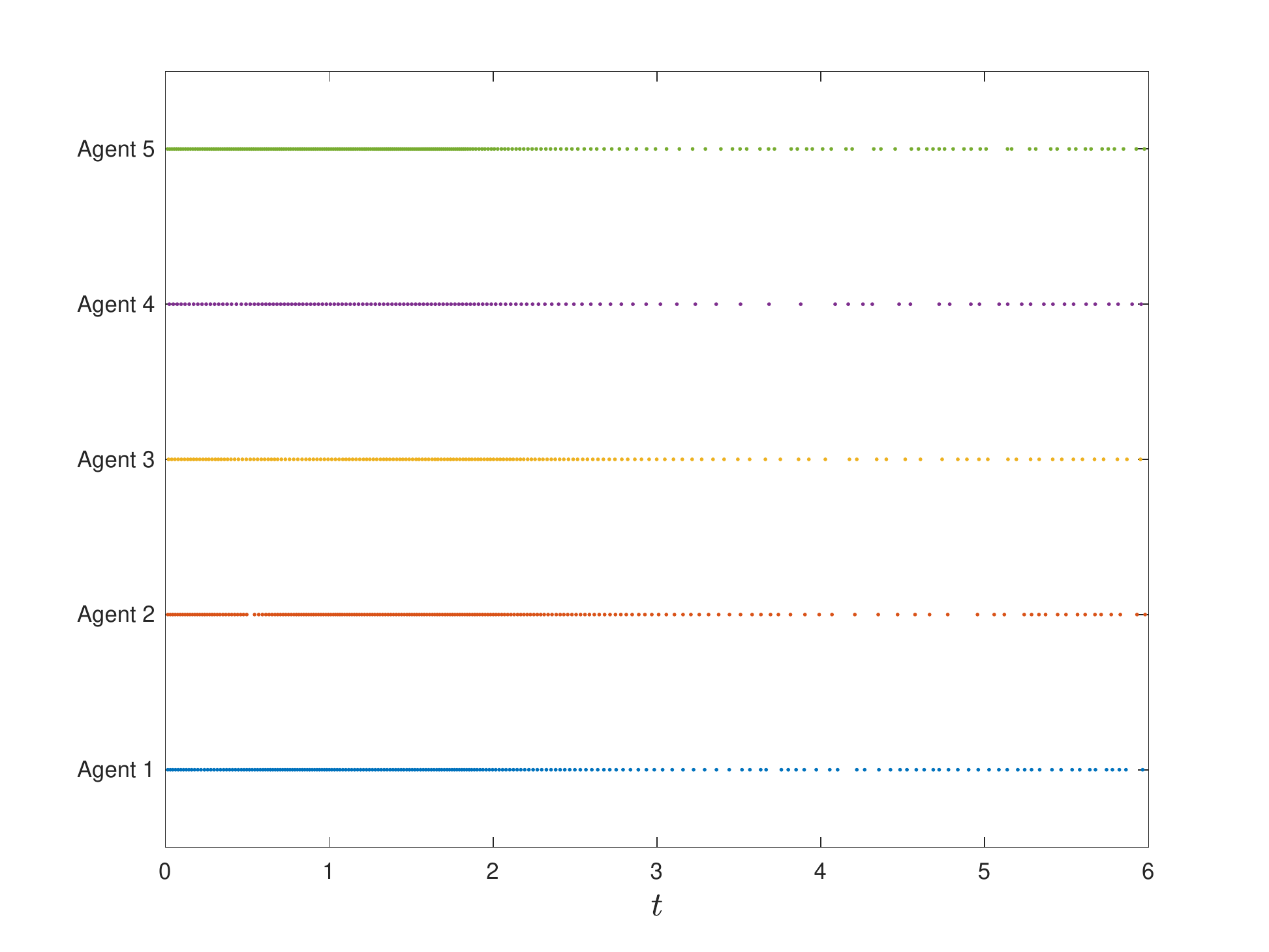}
\par\end{centering}
\caption{The triggering time instants of each agent in the multi-agent system
\eqref{eq:LF-overall-event-trigger} under the leader-follower network
$\mathcal{G}_{1}^{\prime}$ in Figure \ref{fig:LF-network}.}

\label{fig:LF-trigger-time}
\end{figure}
\textcolor{black}{{} Under these parameters, the global bipartite leader-follower
consensus can be achieved as shown in Figure \ref{fig:LF-trajectory}.
Sequences of triggering time for each agent are demonstrated in Figure
\ref{fig:LF-trigger-time}. The dimensions of control protocol for
each agent are illustrated in Figure \ref{fig:LF-trigger-control}. }

\begin{figure}[h]
\begin{centering}
\includegraphics[width=8.5cm]{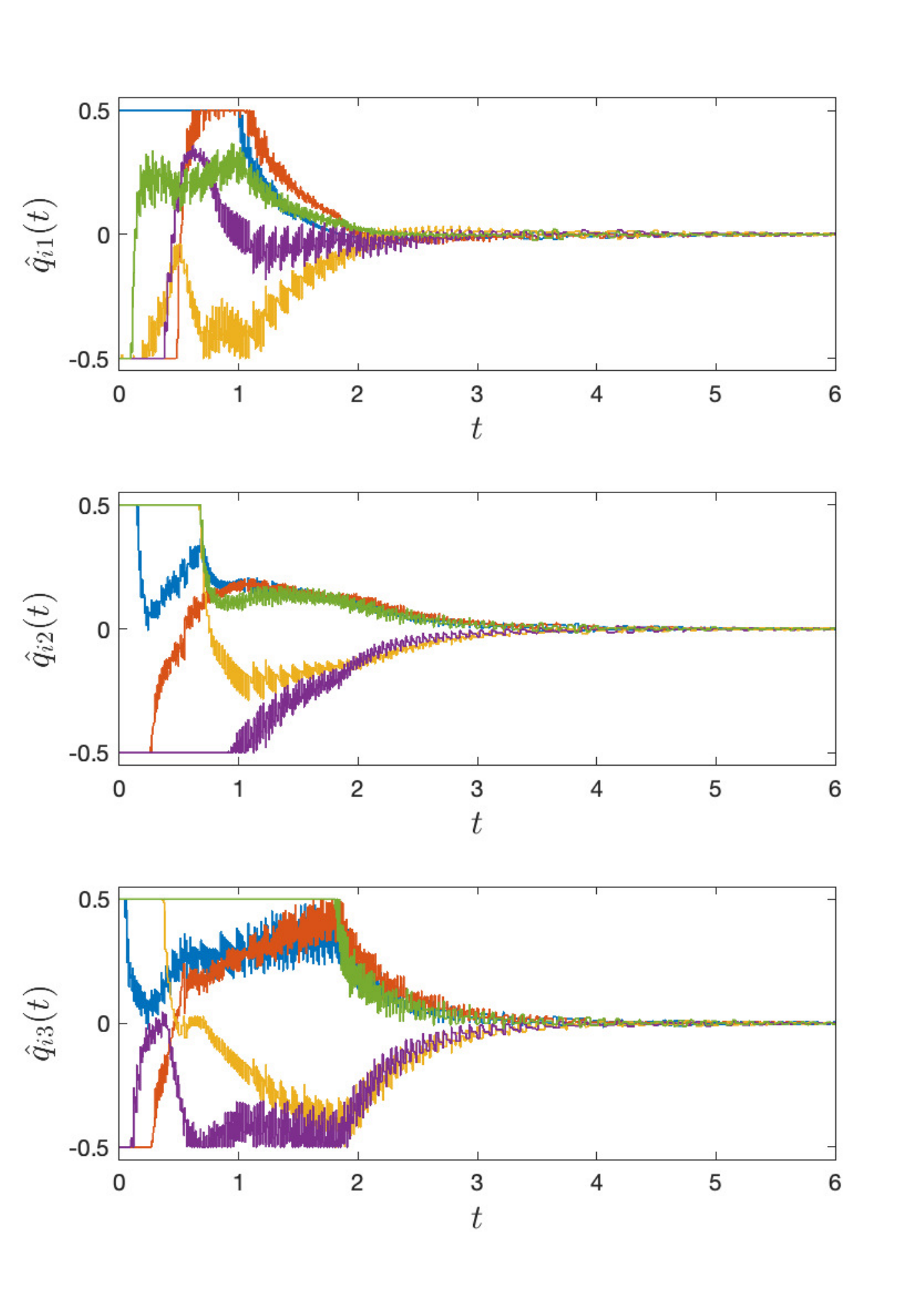}
\par\end{centering}
\caption{The event-based control protocol\textcolor{black}{{} $\widehat{\boldsymbol{q}}_{i}(t)$
}of each agent $i\in\mathcal{V}$ for the multi-agent system \eqref{eq:LF-overall-event-trigger}
under the leader-follower network $\mathcal{G}_{1}^{\prime}$ in Figure
\ref{fig:LF-network}.}

\label{fig:LF-trigger-control}
\end{figure}
\textcolor{black}{}

\section{\textcolor{black}{Conclusion }}

\textcolor{black}{In this paper, we examined the event-triggered global
bipartite consensus problem for multi-agent systems on matrix-weighted
networks subject to input saturation constraints. Dynamic event-triggered
distributed protocols for both leaderless and leader-follower cases
are provided, where each agent only needs to broadcast at its own
state on triggering times, and listen to incoming information from
its neighbors at their triggering times, which reduces the limited
communication resource and avoids the continuous communication among
agents. Then, some criteria are derived to guarantee the leaderless
and leader-follower global bipartite consensus of the multi-agent
systems. Also, the proposed triggering laws are shown to be free of
Zeno phenomenon by proving that the triggering time sequence of each
agent is divergent. Simulation examples demonstrate the effectiveness
of the proposed methods. }

\bibliographystyle{ieeetr}
\phantomsection\addcontentsline{toc}{section}{\refname}\bibliography{mybib,mybib-dynamic-event-trigger}

\end{document}